\documentclass[sigconf,nonacm]{acmart}

\setcopyright{acmcopyright}
\copyrightyear{2023}
\acmYear{2023}
\acmDOI{XXXXXXX.XXXXXXX}

\usepackage{kotex}

\acmSubmissionID{pos-118s1}

\begin{document}

\title{Learning Human-like Locomotion Based on Biological Actuation and Rewards}

\author{Minkwan Kim}
\affiliation{%
  \country{Hanyang University, South Korea}}
\email{palkan21@hanyang.ac.kr}
\author{Yoonsang Lee}
\affiliation{%
  \country{Hanyang University, South Korea}}
\email{yoonsanglee@hanyang.ac.kr}

\renewcommand{\shortauthors}{Minkwan Kim and Yoonsang Lee}

\begin{abstract}
We propose a method of learning a policy for human-like locomotion via deep reinforcement learning based on a human anatomical model, muscle actuation, and biologically inspired rewards, without any inherent control rules or reference motions.
Our main ideas involve providing a dense reward using metabolic energy consumption at every step during the initial stages of learning and then transitioning to a sparse reward as learning progresses,
and adjusting the initial posture of the human model to facilitate the exploration of locomotion.
Additionally, we compared and analyzed differences in learning outcomes across various settings other than the proposed method.

\end{abstract}

\maketitle

\section{Introduction}
In recent years, there has been significant progress in research on learning control polices for physically-simulated characters via deep reinforcement learning. 
In most studies, the torque calculated by the target pose output by the policy or the torque directly output by the policy is applied to each joint \cite{yu_learning_2018, kang_finite_2021}
and a policy learns to produce motion results similar to reference motions using an imitation reward \cite{peng_deepmimic_2018}.
However, while many successful results have been reported, this approach has limitations in that it differs from the way actual humans move their joints by contracting muscles, and the generated motion is influenced by the reference motion used.

To produce more human-like movements, there have been consistent attempts to control characters actuated by musculotendon units rather than joint torque.
Wang et al. proposed a walking control method for a simple model consisting of 16 muscles, assuming control rules based on swing / stance phases and optimizing the parameters used in the rule, which provided some prior information on ``walking''~\cite{wang_optimizing_2012}.
Lee et al. proposed a method to control the model consisting of over one hundred muscles using quadratic programming and trajectory optimization, but has a limitation that range of generated result motion is restrained to reference motions used~\cite{lee_locomotion_2014}.
\cite{lee_scalable_2019} proposed a method for training a policy to reproduce a given reference motion for a model with a larger number of muscles than former, using deep reinforcement learning.

In contrast to these existing approaches, we propose a method to train a policy for human-like locomotion through deep reinforcement learning based on an anatomical humanoid model, musculotendon actuators, and biological rewards including metabolic energy consumption.
Because no reference motions or feedback control rules are used, the policy is trained solely to maximize cumulative rewards without any bias.
However, unlike existing approaches that use imitation rewards which provides more direct guidance for good actions during the early stages of learning, this approach is prone to getting stuck in local minima during the early stages.
We have found that a reward for metabolic energy minimization actually helps learning in such situations, contrary to common belief.
This can be interpreted as the energy reward, similar to the imitation reward, providing more direct information about ``good'' movements to facilitate learning.
This supports the well-known belief that human natural walking is the type of locomotion that expends the least amount of energy.
We showed that by changing the dense energy reward to a slightly modified form of sparse reward after a certain amount of learning, the policy can be trained to perform stable human-like locomotion.
We also found that the initial posture of the musculoskeletal model at the beginning of each episode plays an important role in the learning process.

\section{Simulation and Learning}
We use the Hill-type muscle model where %
a musculotendon unit is composed of a muscle fiber and tendon.
To improve computational speed, we assumes a fixed length for the tendon at its resting state while simulating the musculotendon units, as in many previous studies.
A muscle fiber consists of an active element that actively generates contractile force and a passive element that models the elastic force that is generated passively when the muscle is stretched beyond a certain length. The overall musculotendon force is calculated as follows~\cite{zajac_muscle_1989}:
\begin{equation}
  f_{mtu} = \cos{\alpha} \cdot (a \cdot g_{al}(l_m) \cdot g_{v}(\dot{l}_m) + g_p(l_m)),
\end{equation} %
where $l_m$ and $\dot{l}_m$ for muscle fiber length and its time derivative, $\alpha$ for pennation angle, and $a$ for muscle activation.
$g_{al}$ and $g_v$ represent force-length and force-velocity functions of active element, respectively, and $g_p$ represent the force-length function of passive element.
We adopted the same functions as those in \cite{lee_locomotion_2014}.

Our policy receives positions and velocities of joints and links, and muscle fiber lengths as 278-dimensions input, and outputs 120-dimensions muscle activation.
The reward function is as follows:
\begin{equation}
    r(s, a) = r_{up} + r_{ori} + r_{eng} + r_{vel} + r_{dev} + r_{alive}.
\end{equation}
$r_{up}$, $r_{ori}$, and $r_{eng}$ are biological rewards that reflect the characteristics of human locomotion, and correspond to terms that keep the upper body straight, prevent pelvis from tilting, and minimize energy consumption, respectively.
$r_{vel}$, $r_{dev}$, and $r_{alive}$ are goal rewards that encourage the model to achieve a desired velocity, maintain the center of mass in a straight line without deviating to the sides, and avoid falling over for as long as possible, respectively.
Each reward term includes a weight to ensure that they are balanced to similar magnitudes in the overall reward function.

\section{Design for Human-like Locomotion}

\paragraph{Energy Rewards.} Our energy reward is based on the metabolic energy consumption of muscles, which is calculated using factors such as muscle activation, volume, and fiber velocity \cite{wang_optimizing_2012}.
We use two types of energy reward as follows:
\begin{itemize}
    \item Metabolic Equivalent of Task (MET): MET is the metabolic energy consumption rate normalized by body weight. This is calculated at each step, as a dense reward.
    \item Cost of Transport (CoT): CoT is the quantity of energy required to move a unit distance. %
        This is calculated only once at the end of each episode, as a sparse reward. 
\end{itemize}
We employ a two-stage training approach to achieve stable human-like locomotion in our policy. Specifically, we first train the policy using the dense reward (MET) until convergence, and then switch to the sparse reward (CoT) to further refine and stabilize the learned policy.
In the early stages of learning, the policy's activations often result in movements with high energy consumption and instability.
The dense reward (MET) aids in rapidly stabilizing the movements, facilitating the discovery of a policy that can maintain balance over multiple steps.
Afterward, switching to the sparse reward (CoT) encourages covering longer distances even with the same energy consumption, effectively increasing the travel distance.

\paragraph{Initial Posture.} We found that the humanoid's posture at the beginning of each episode had a significant impact on whether or not it could learn human-like locomotion.
If training begins with both feet in contact, the policy is learned to repeatedly jump with both feet.
We speculate that this is because two-legged jumping actions are more likely to be explored than one leg lifting in the early stages.
The policy successfully learned human walking behavior by introducing a randomized starting position of lifting either the left or right leg at the beginning of each episode.
This approach increased the probability of exploring actions that naturally led the lifted leg to make contact with the ground, while also enabling the policy to learn how to swing the opposite leg.
As a result, the policy acquired the ability to swing both legs alternately, ultimately enabling it to walk in a manner akin to human locomotion.

\section{Results}
The model has a mass of 75kg, consists of 16 links and 31 DOFs with 120 muscles.
DART and RLlib (with Proximal Policy Optimization) were used for simulation and DRL.
Our training took approximately 10 days to learn a policy for stable walking.
For detailed results, refer to the supplemental document and accompanying video.

\paragraph{Dense and Sparse Energy Rewards.}
With the dense reward (MET) only, the policy converged after taking only four to five walking steps.
With the sparse reward (CoT) only, the policy failed to learn how to move forward and fell easily. 
We could train stable, human-like locomotion by sequentially using two types of energy rewards over two stages (Row 1--3 of Figure 1).

\paragraph{Importance of Energy Rewards.}
When no energy reward was used, the policy did not learn how to move at all, resulting in the humanoid falling over immediately.
We speculate that, in the case of musculoskeletal models, minimizing the energy consumption is not just about limiting the average force magnitude, but also about giving advantage to explore state and action spaces that are closer to human-like movements (Row 4 of Figure 1).

\paragraph{Importance of Initial Posture.}
As previously discussed, starting training with both feet in contact resulted in the learned policy involving repeated jumping with both feet, while introducing a randomized starting position of lifting either the left or right leg at the beginning of each episode was successful in teaching the policy human-like locomotion behavior (Row 5 of Figure 1).

\paragraph{Ablation Study.}
The humanoid could not move forward properly by a policy that learned with the mean squared sum of muscle activations as the energy reward.
When excluding 120-dimensional muscle fiber lengths from our observation, the total reward increased slowly and the humanoid could not move forward properly as well, resulted in a quick fall  (Row 6--7 of Figure 1).

\section{Conclusions and Future Works}
We demonstrated a new method for learning policies to generate human-like locomotion of muscle-actuated humanoids, which requires no inherent control rules or reference motions.
The key idea is the sequential adoption of dense and sparse metabolic energy rewards in two stages, along with an initial posture that facilitates alternating leg swings.
The current results have some differences from actual human movements, such as walking without swinging the arms or the front part of the foot touching the ground first. We plan to improve these aspects and expand our research to cover a wider range of movements, such as running and jumping.

\begin{acks}
This work was supported by the National Research Foundation of Korea (NRF) grant funded by the Korea Government (MSIT) (NRF-2019R1C1C1006778, RS-2023-00222776).
\end{acks}

\bibliographystyle{ACM-Reference-Format}
\bibliography{sample-base}

\newpage

\begin{figure*}[h]
  \centering

  \includegraphics[width=0.1\linewidth]{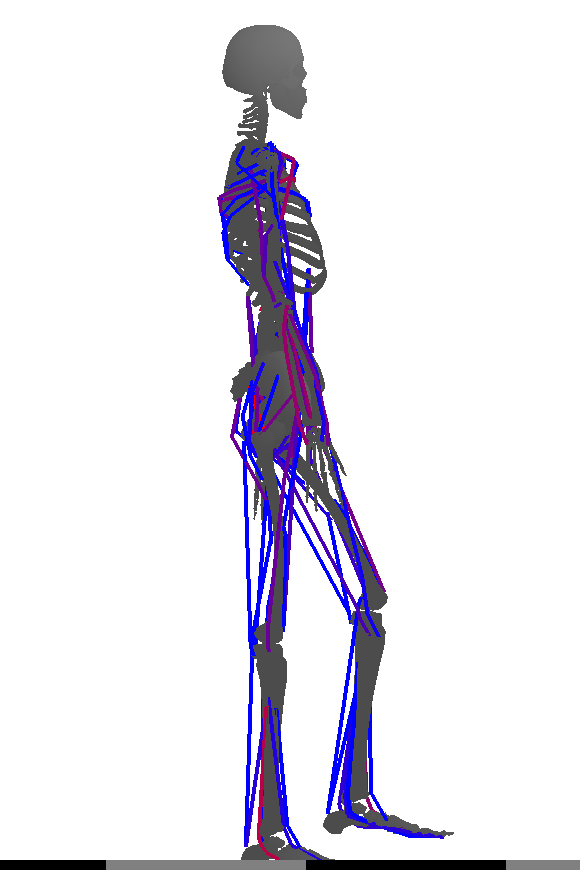}
  \includegraphics[width=0.1\linewidth]{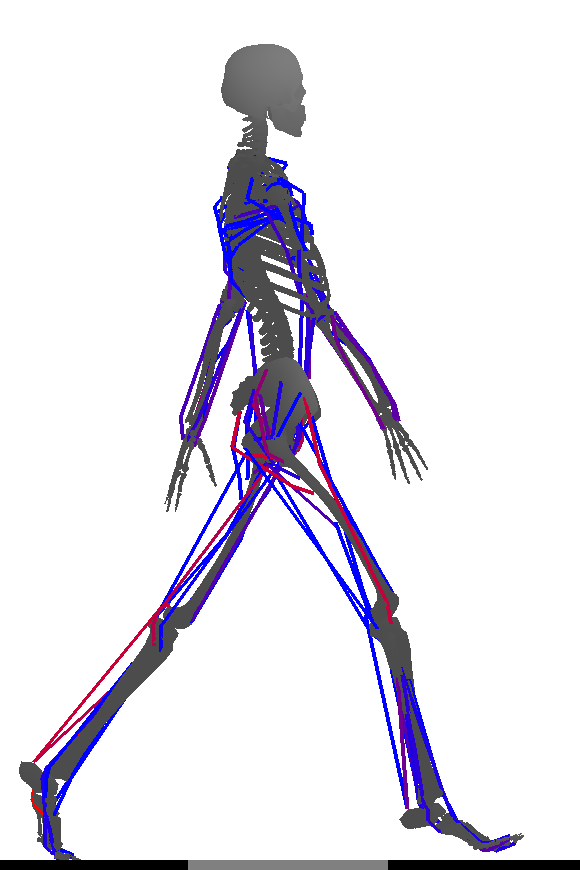} 
  \includegraphics[width=0.1\linewidth]{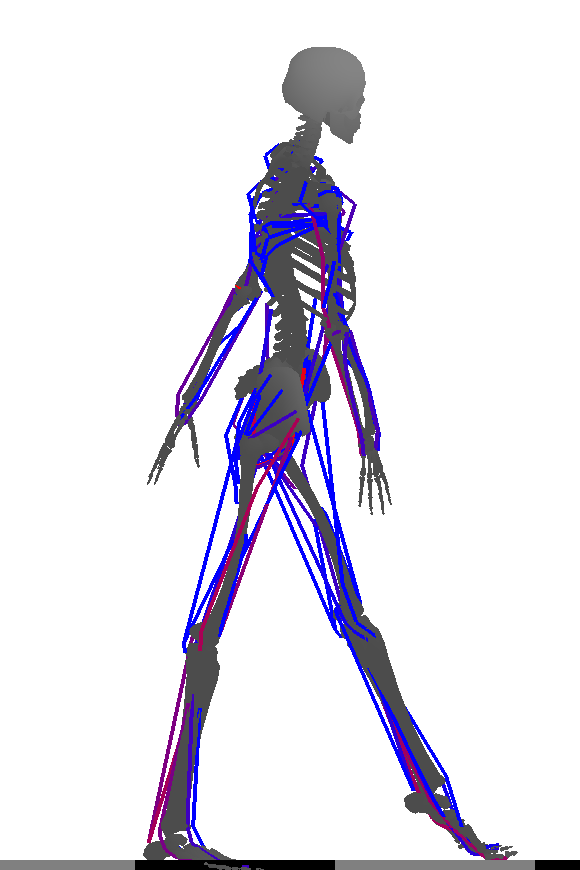}
  \includegraphics[width=0.1\linewidth]{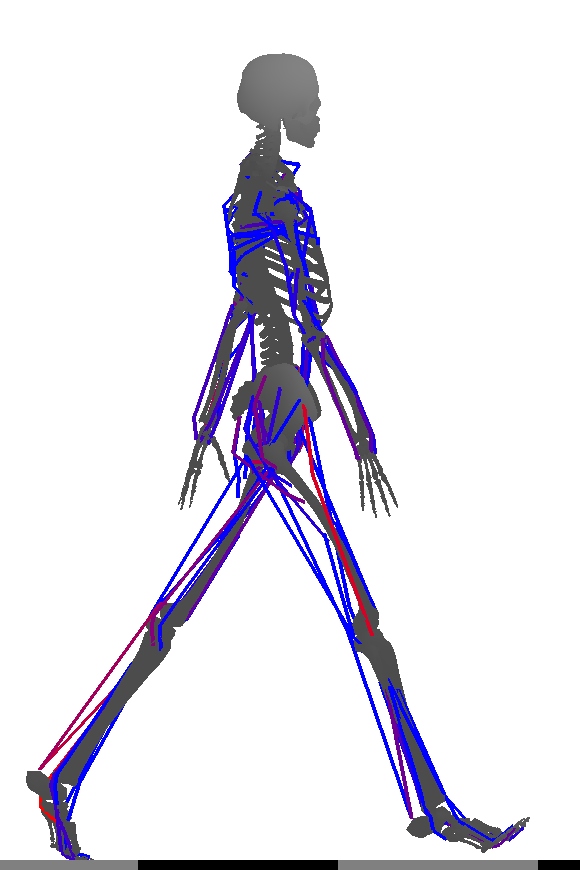}
  \includegraphics[width=0.1\linewidth]{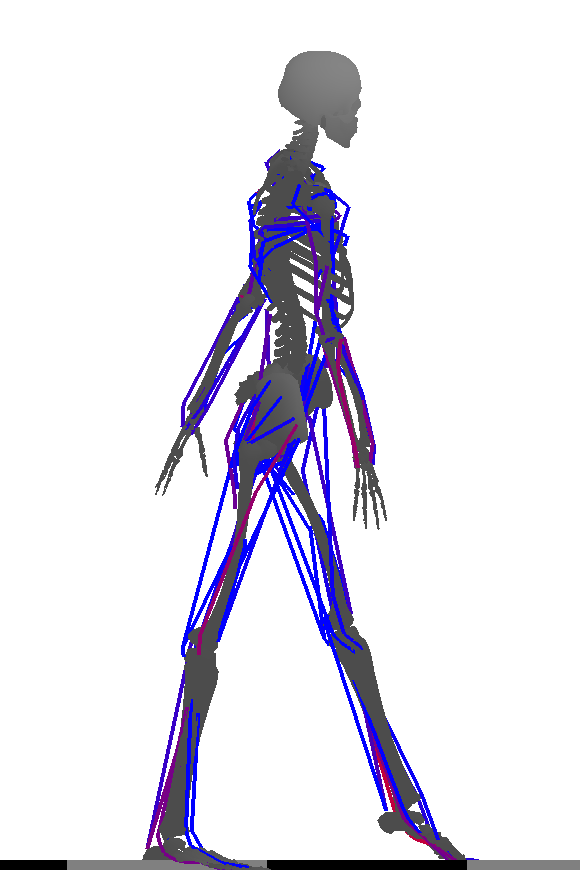} 
  \includegraphics[width=0.1\linewidth]{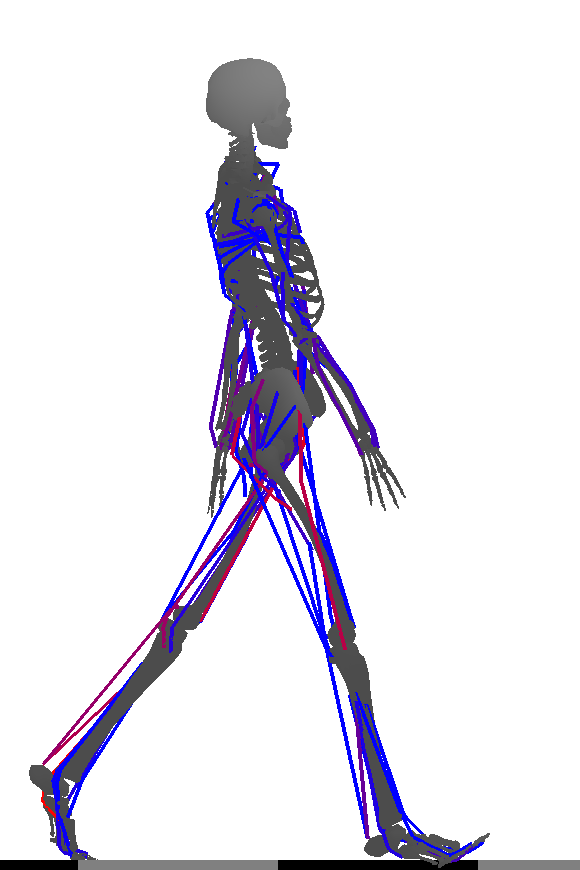} 
  \includegraphics[width=0.1\linewidth]{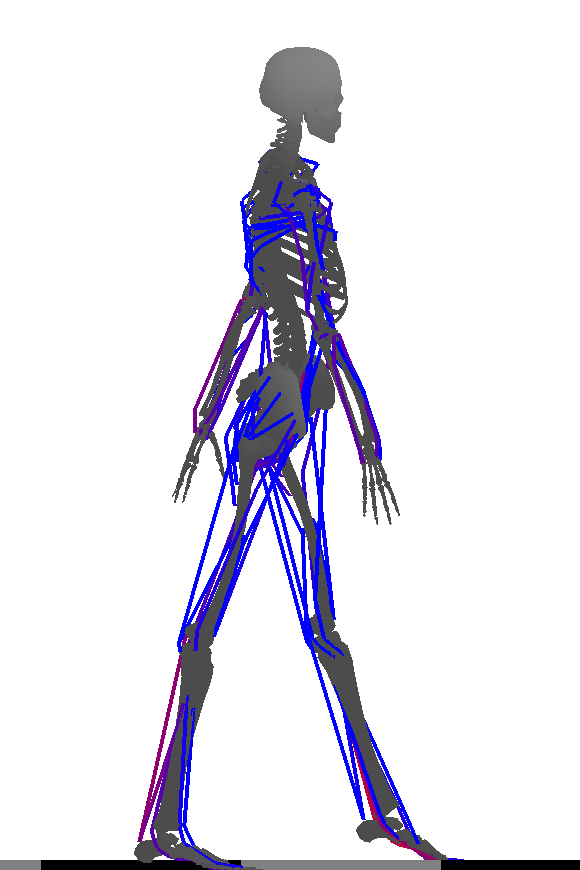}
  \includegraphics[width=0.1\linewidth]{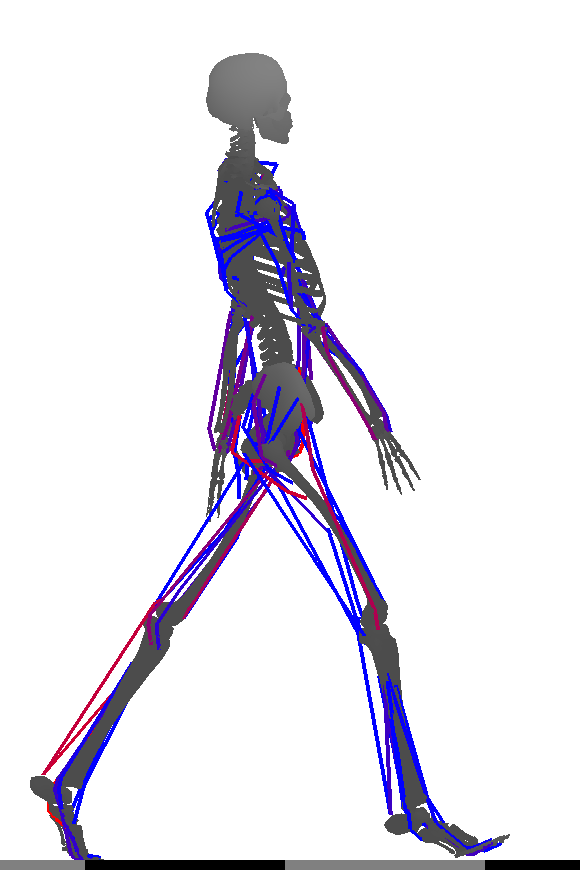} 
  \includegraphics[width=0.1\linewidth]{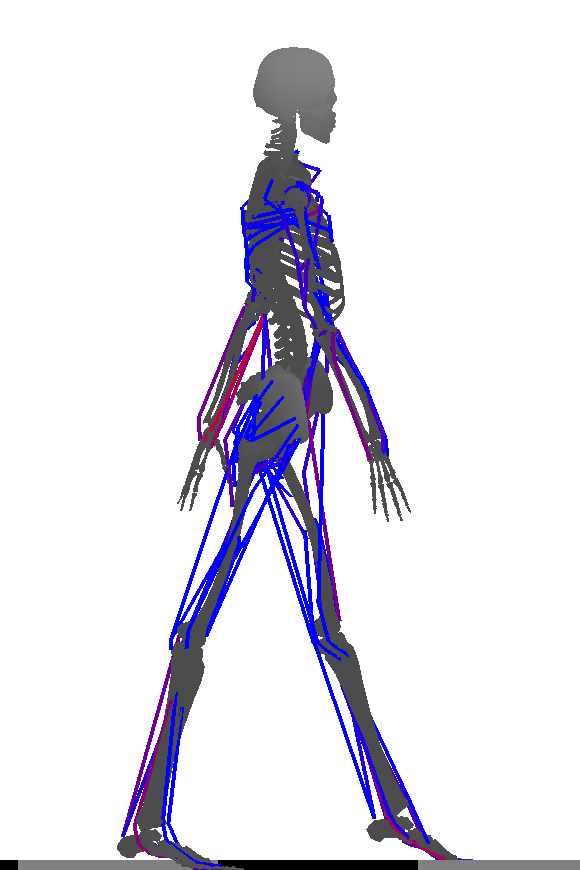}  
  
  \includegraphics[width=0.1\linewidth]{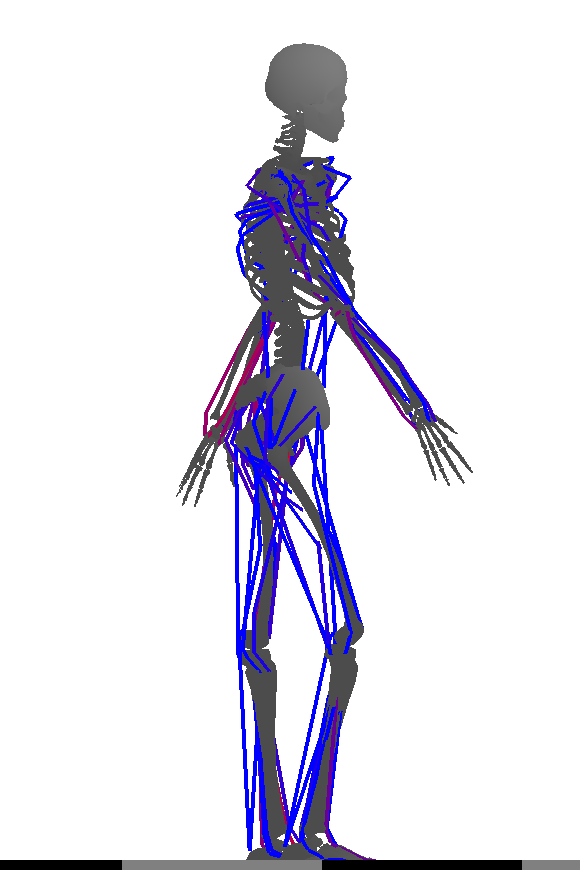}
  \includegraphics[width=0.1\linewidth]{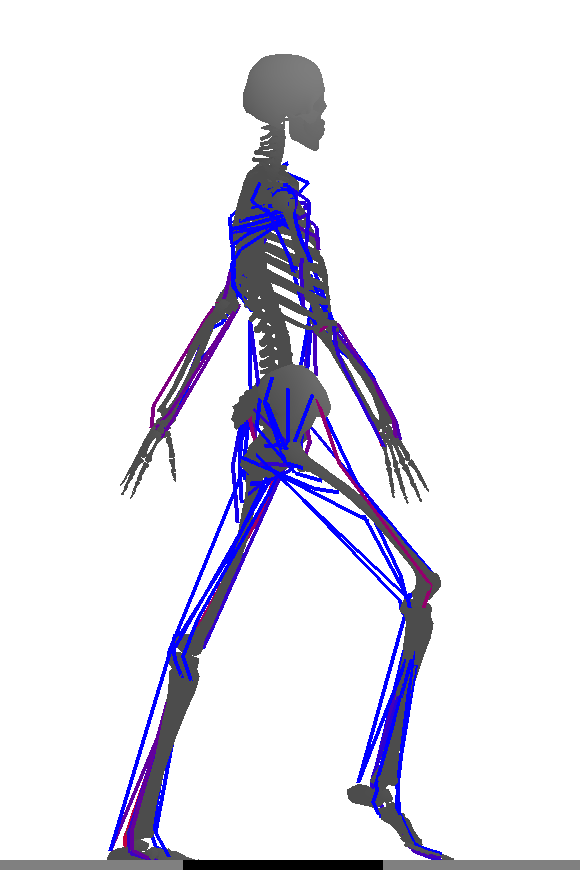} 
  \includegraphics[width=0.1\linewidth]{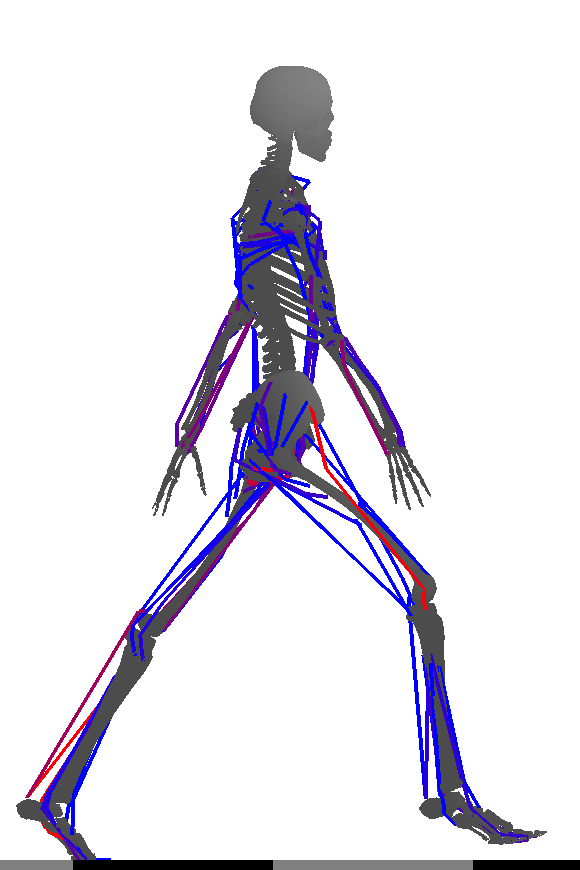}
  \includegraphics[width=0.1\linewidth]{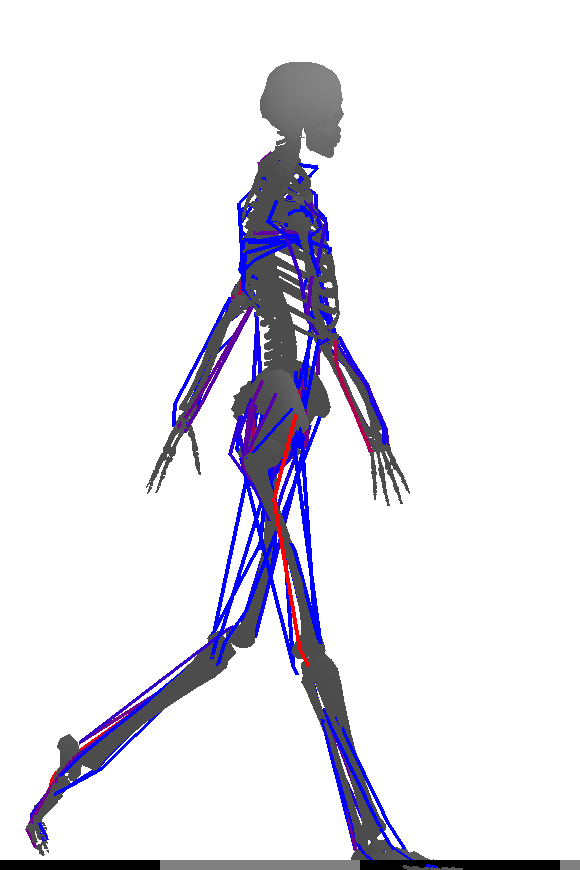}
  \includegraphics[width=0.1\linewidth]{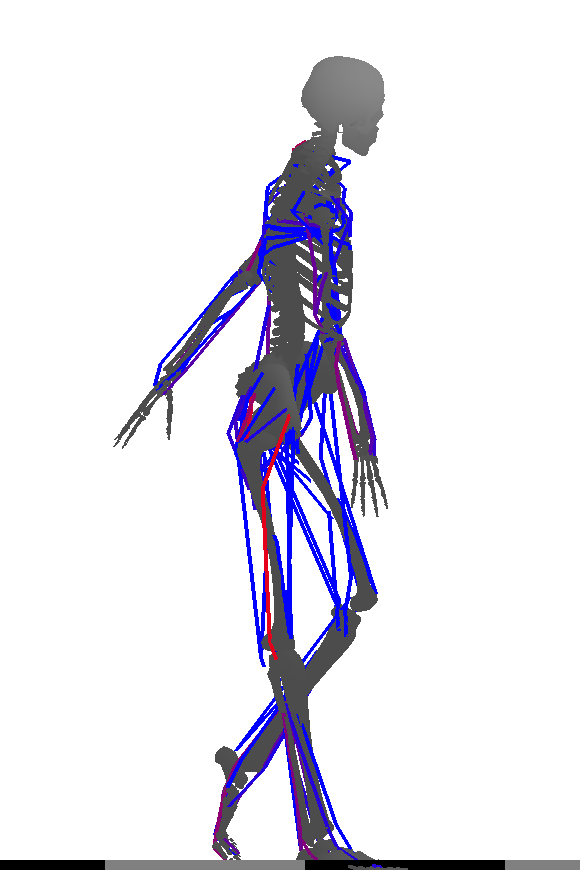} 
  \includegraphics[width=0.1\linewidth]{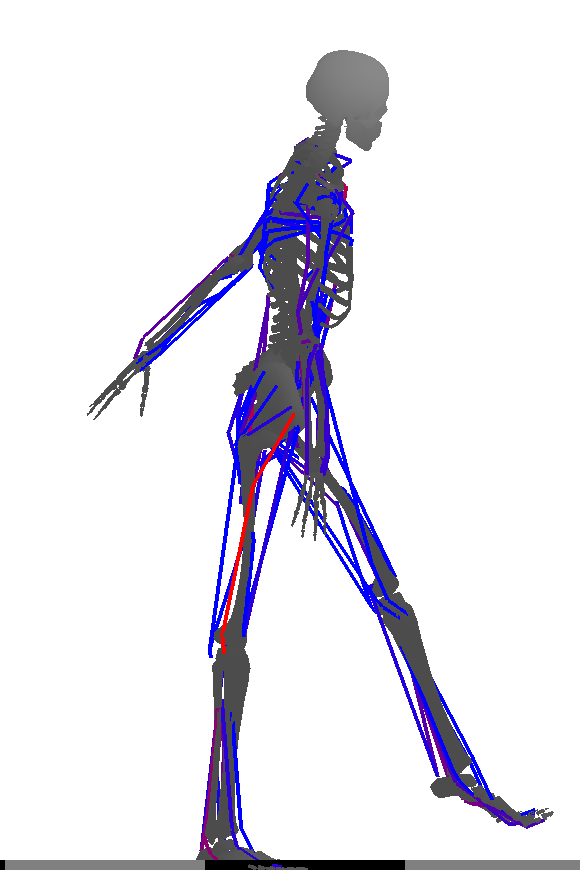} 
  \includegraphics[width=0.1\linewidth]{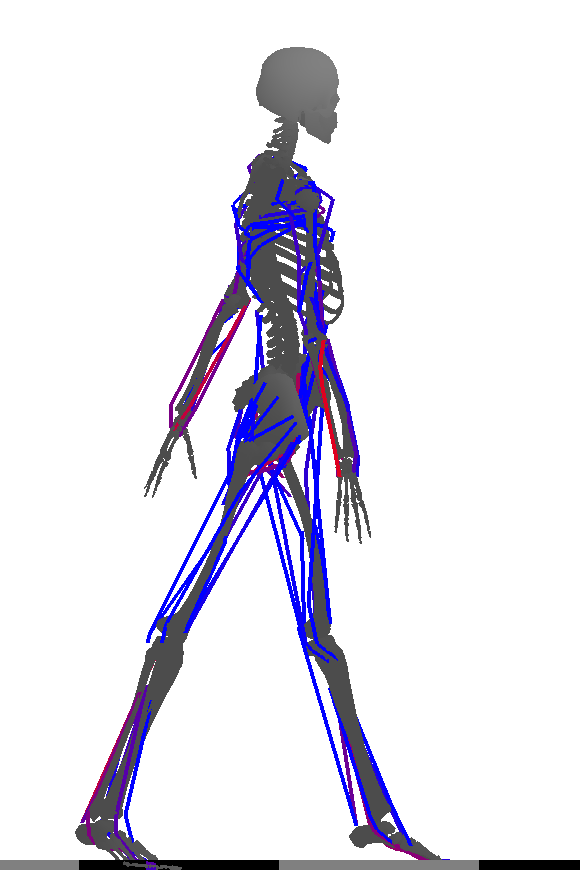}
  \includegraphics[width=0.1\linewidth]{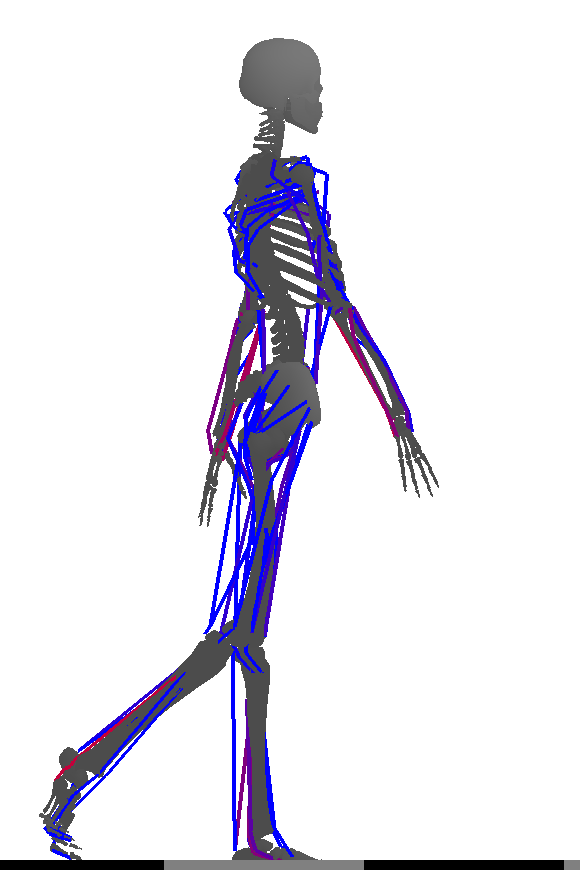} 
  \includegraphics[width=0.1\linewidth]{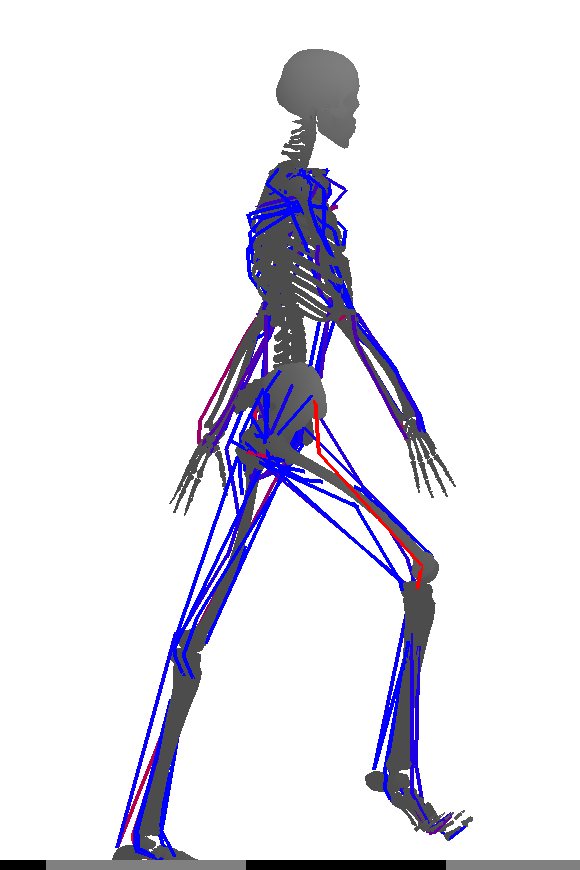}

  \includegraphics[width=0.1\linewidth]{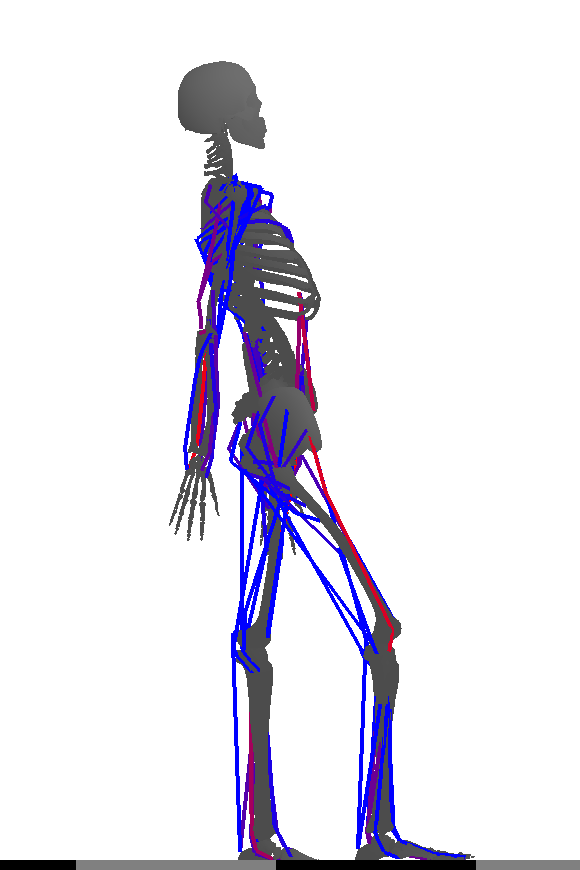}
  \includegraphics[width=0.1\linewidth]{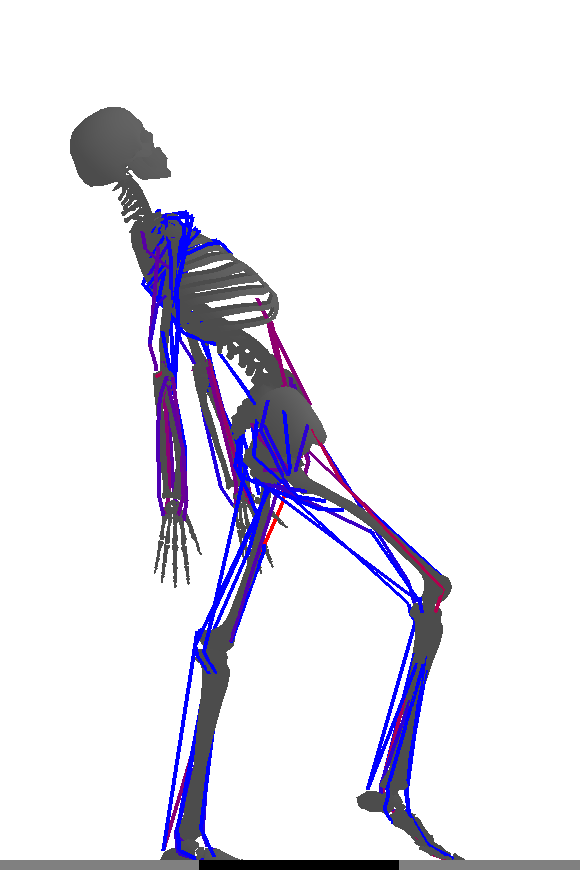} 
  \includegraphics[width=0.1\linewidth]{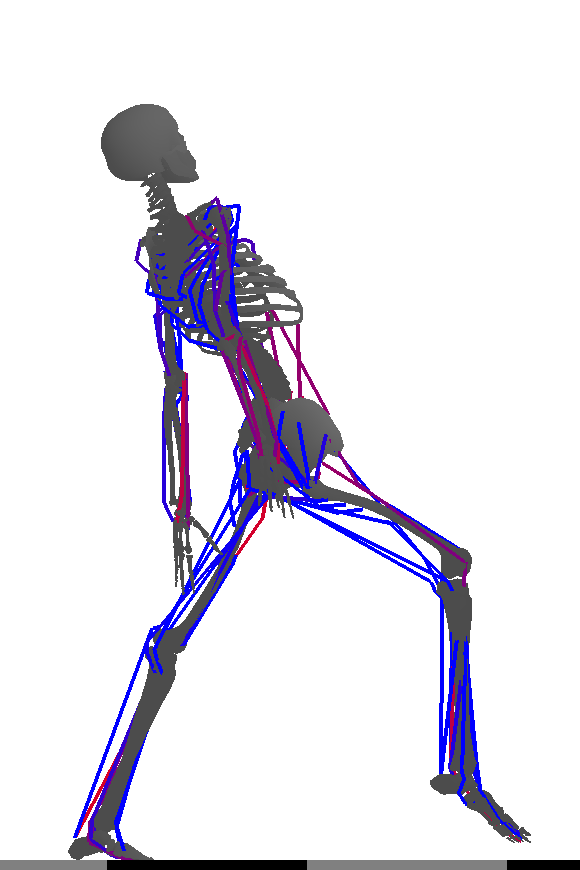}
  \includegraphics[width=0.1\linewidth]{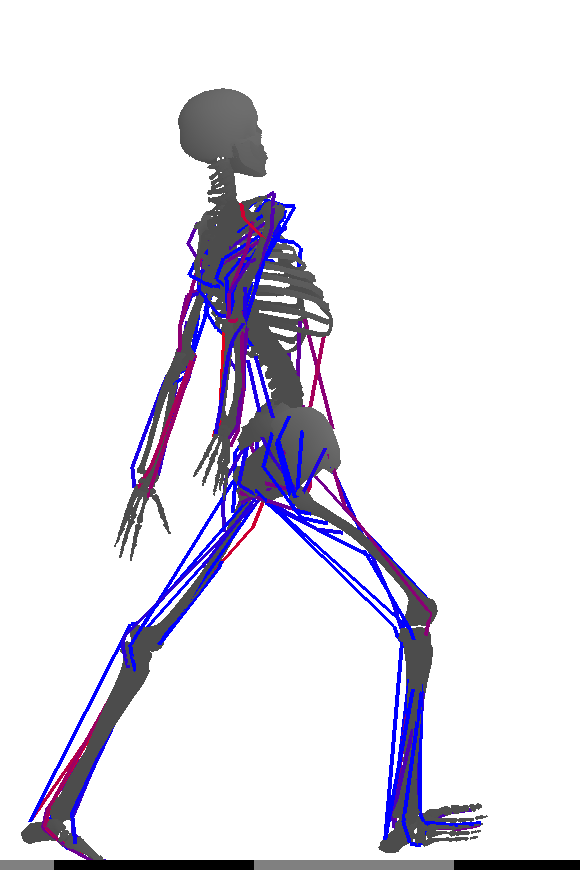}
  \includegraphics[width=0.1\linewidth]{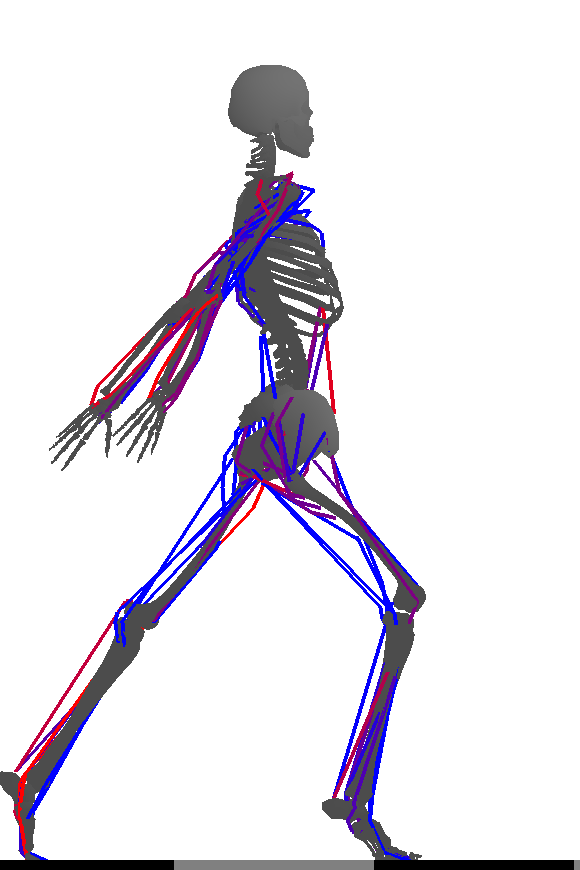} 
  \includegraphics[width=0.1\linewidth]{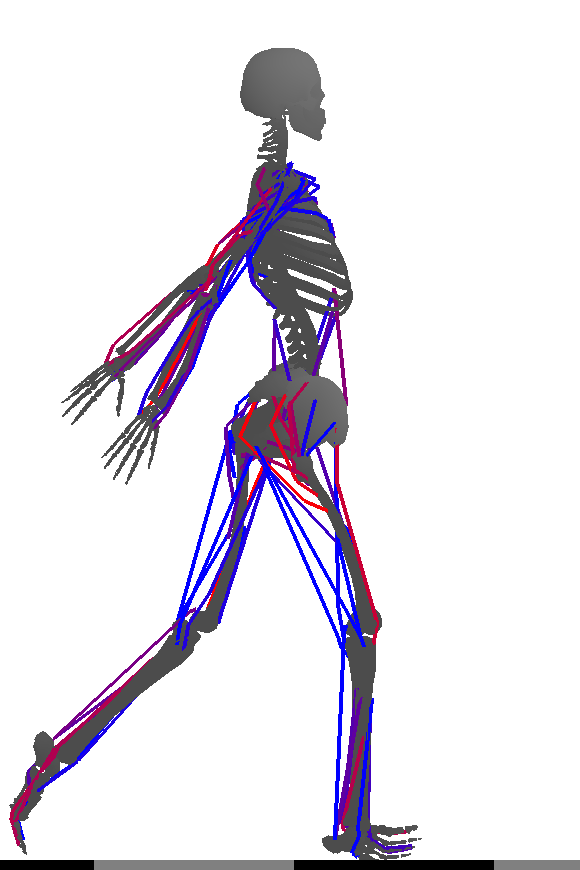} 
  \includegraphics[width=0.1\linewidth]{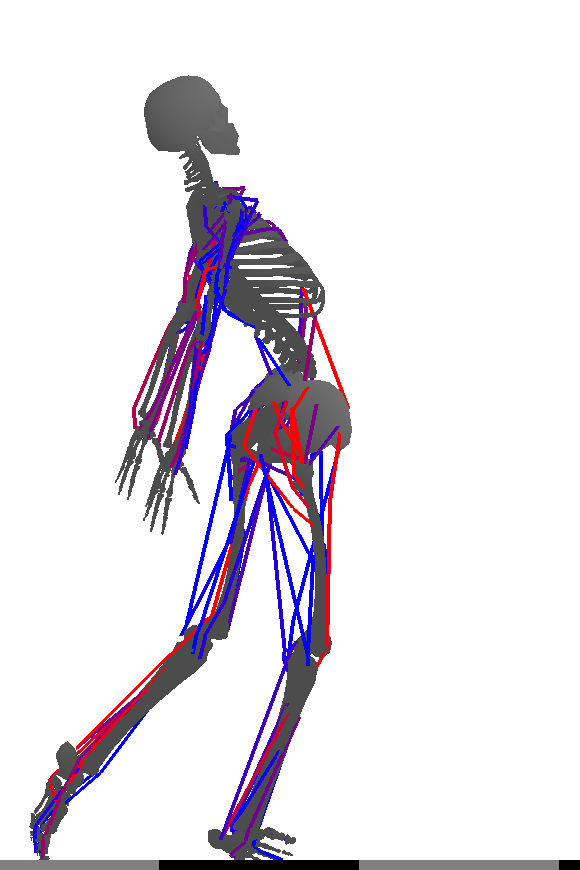}
  \includegraphics[width=0.1\linewidth]{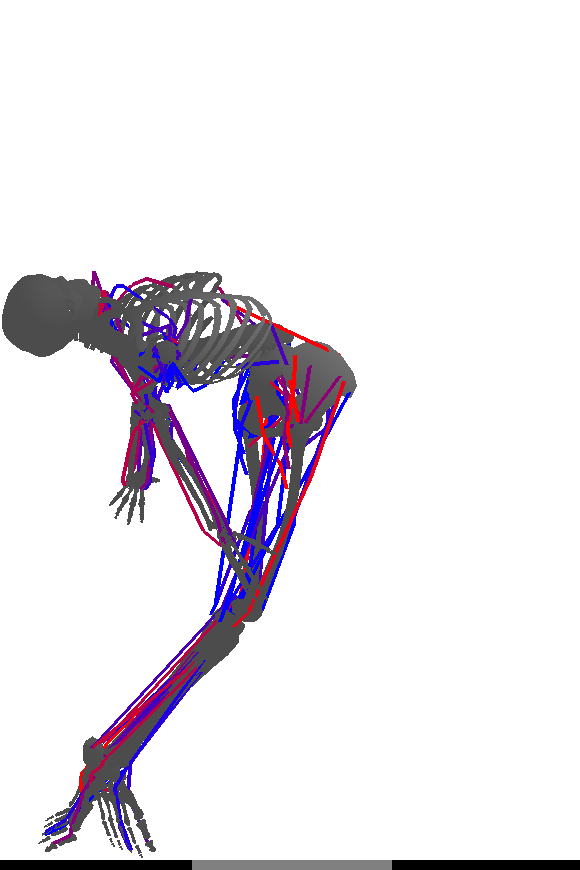} 
  \includegraphics[width=0.1\linewidth]{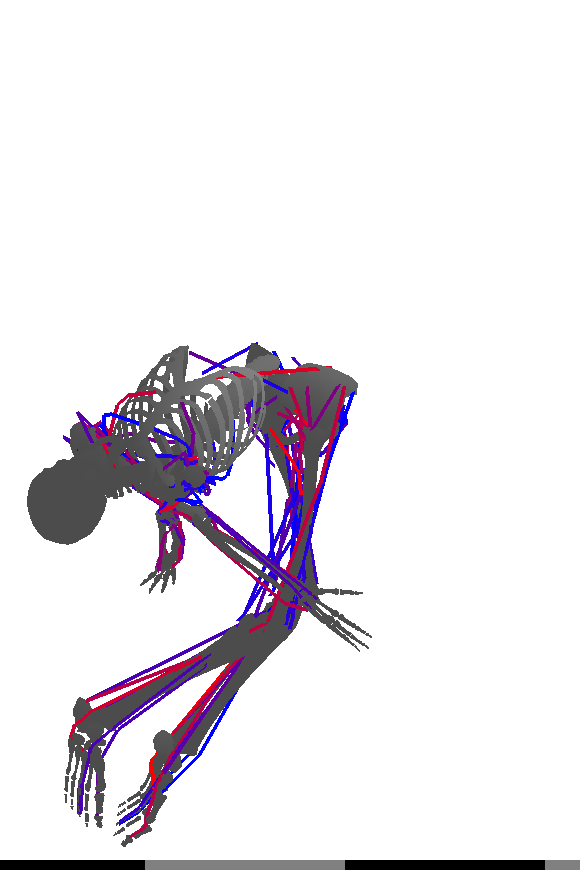}
  
  \includegraphics[width=0.1\linewidth]{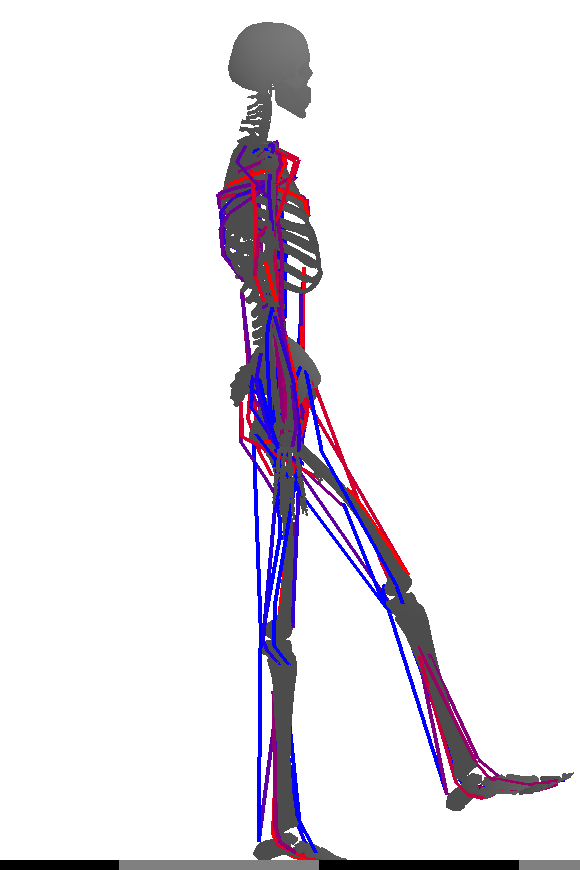}
  \includegraphics[width=0.1\linewidth]{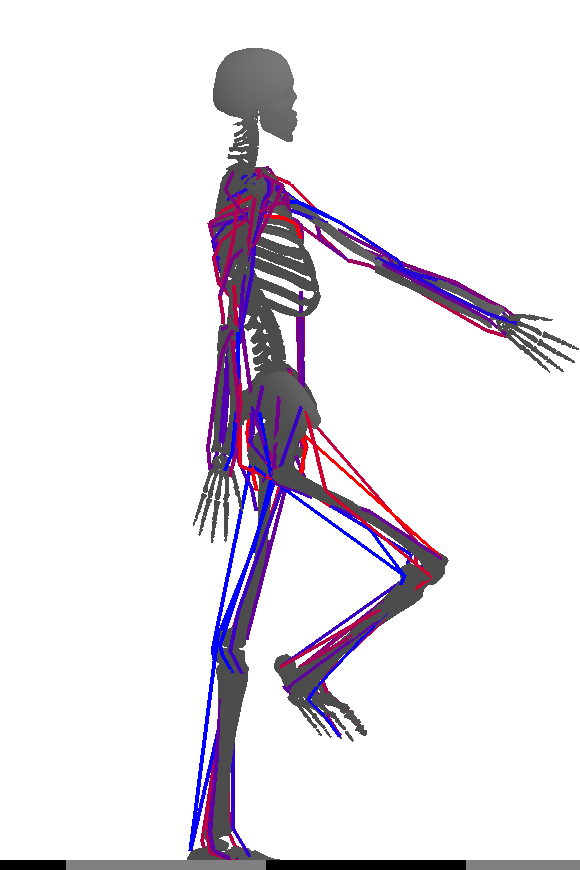} 
  \includegraphics[width=0.1\linewidth]{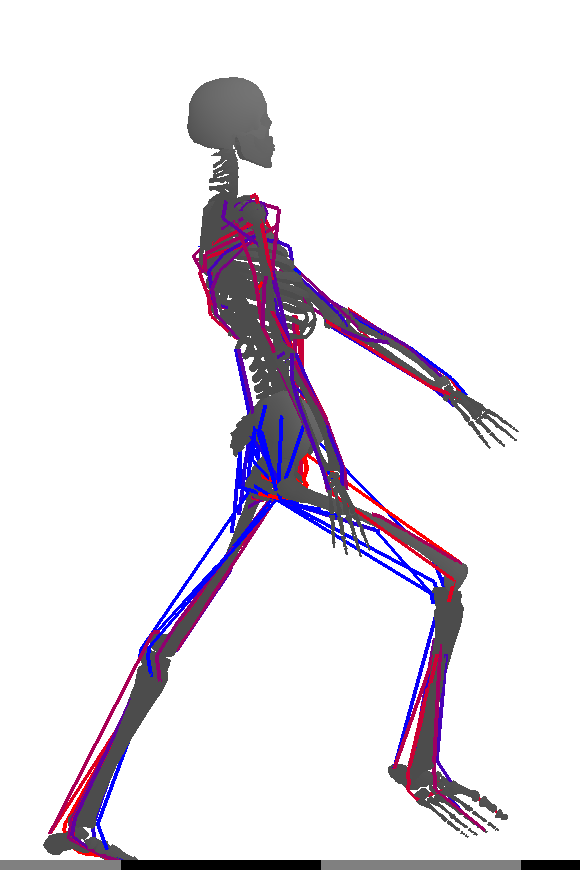}
  \includegraphics[width=0.1\linewidth]{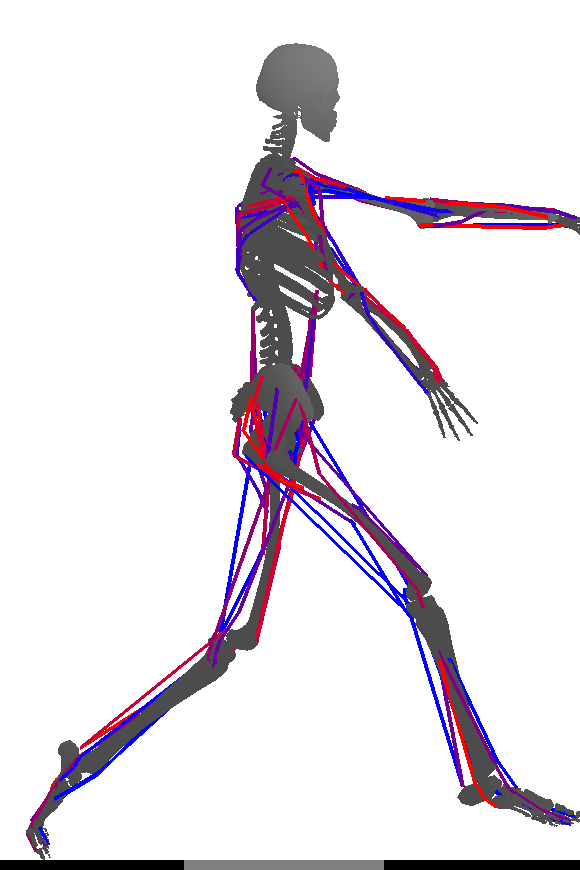}
  \includegraphics[width=0.1\linewidth]{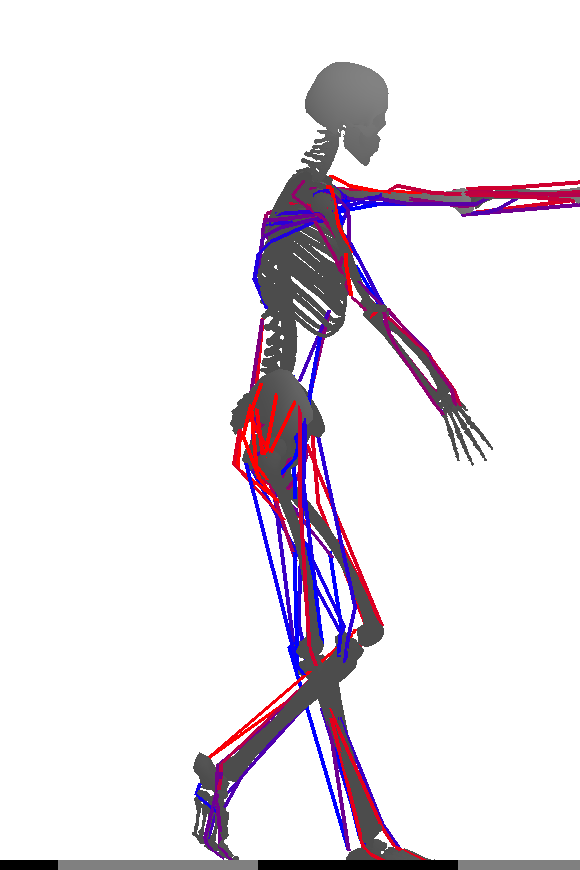} 
  \includegraphics[width=0.1\linewidth]{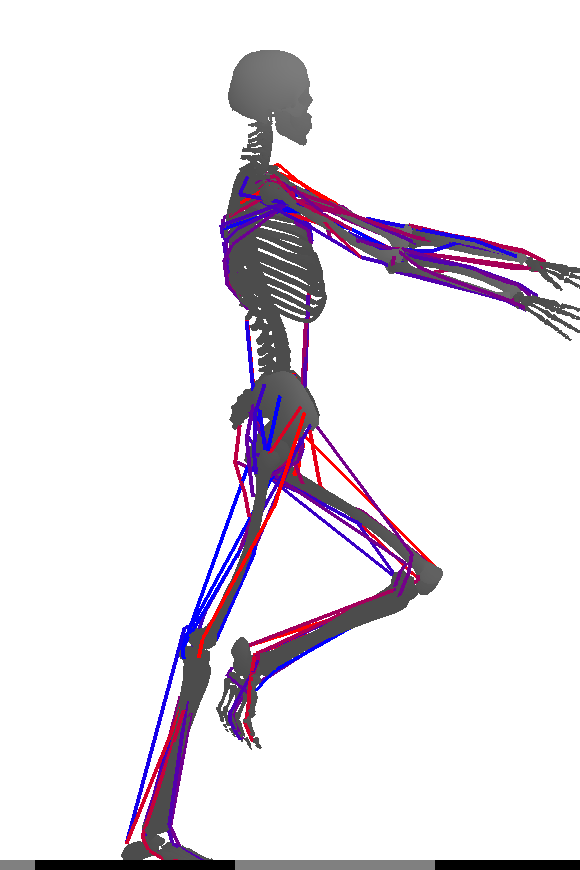} 
  \includegraphics[width=0.1\linewidth]{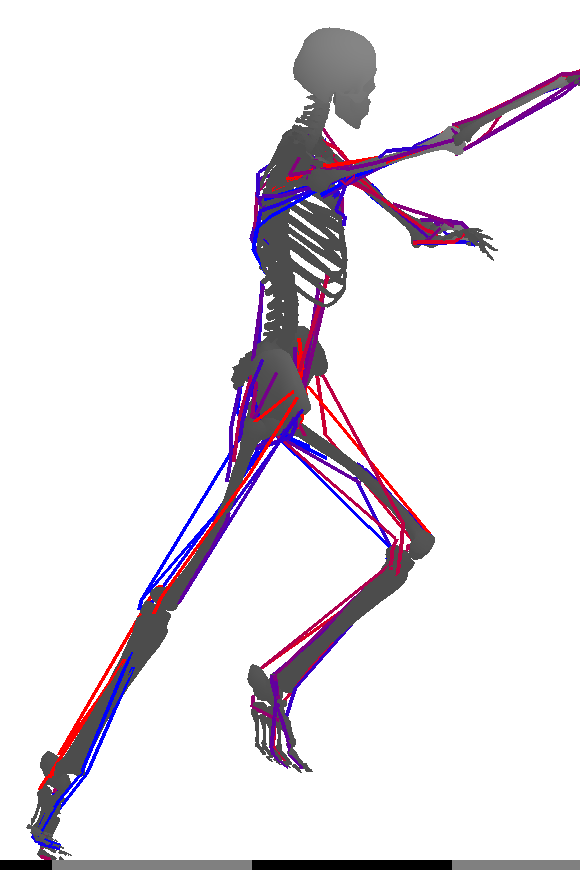}
  \includegraphics[width=0.1\linewidth]{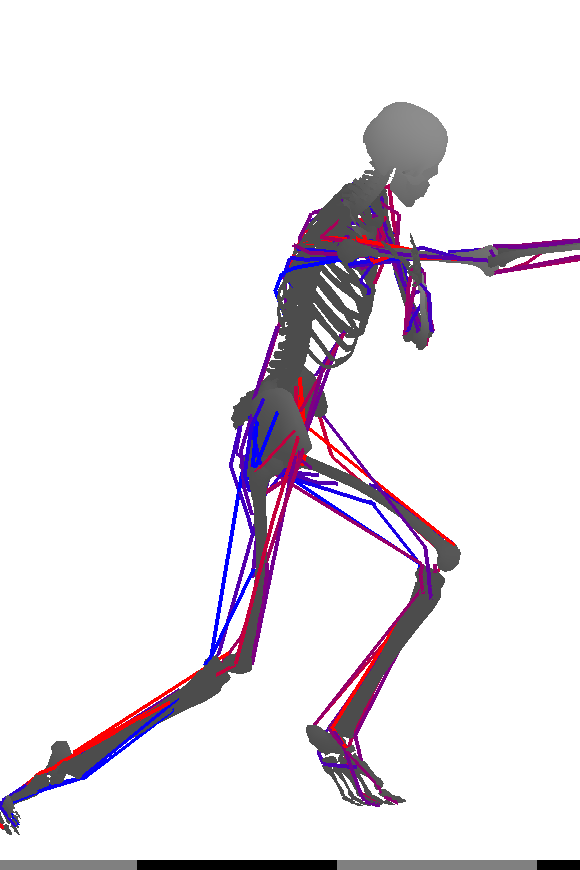} 
  \includegraphics[width=0.1\linewidth]{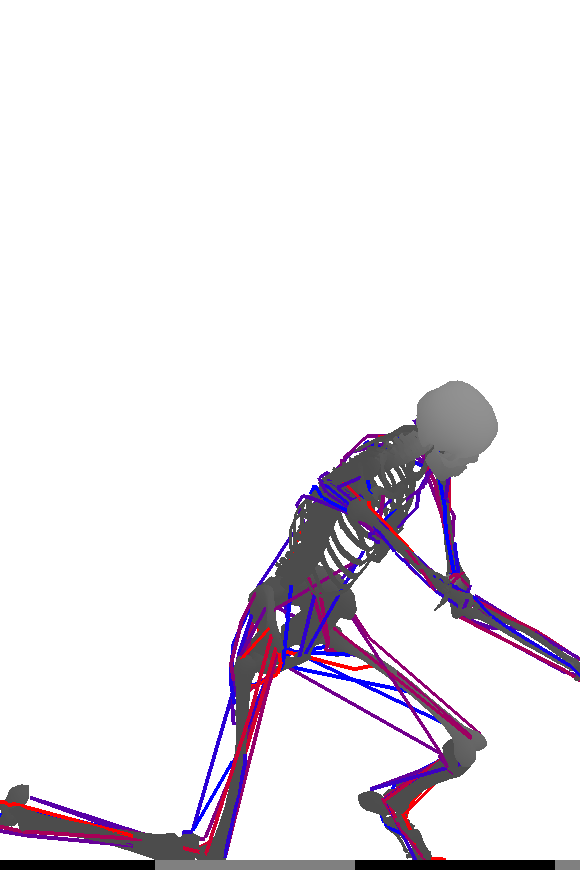} 

  \includegraphics[width=0.1\linewidth]{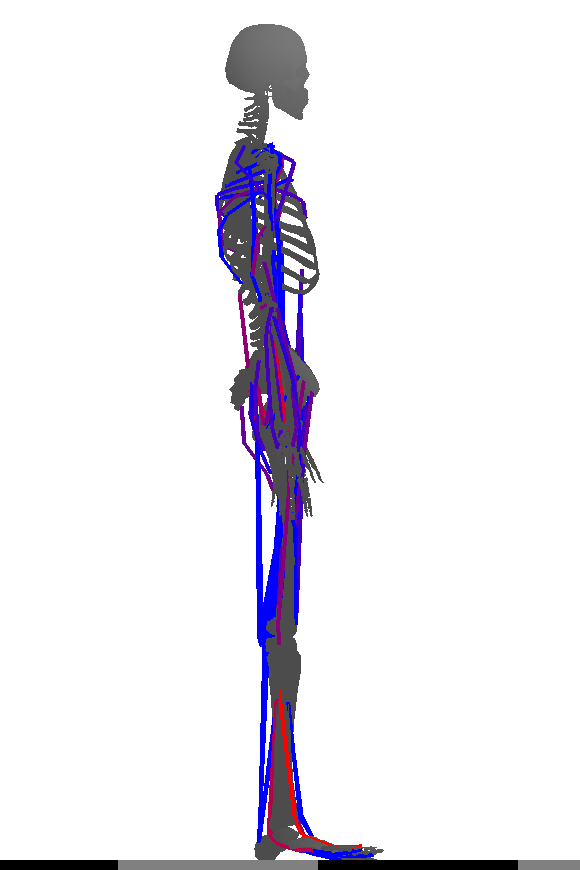}
  \includegraphics[width=0.1\linewidth]{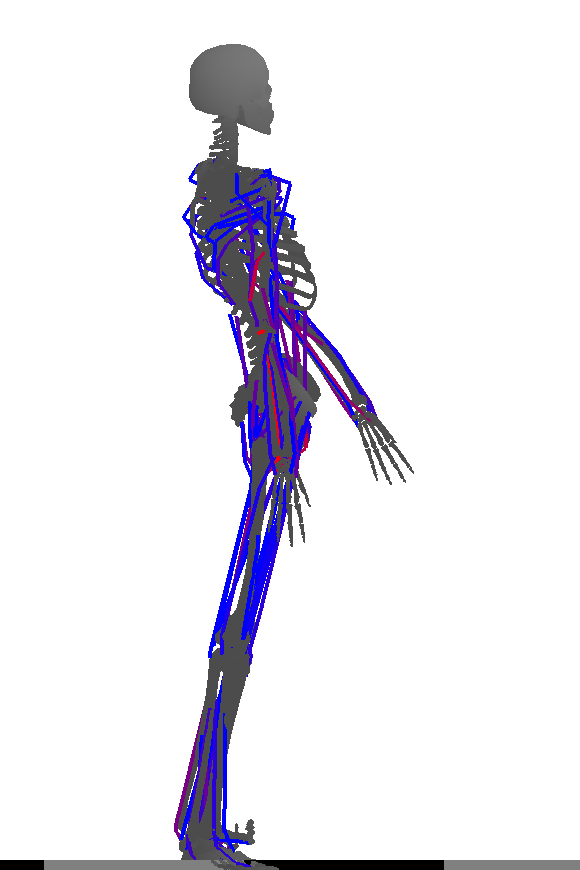} 
  \includegraphics[width=0.1\linewidth]{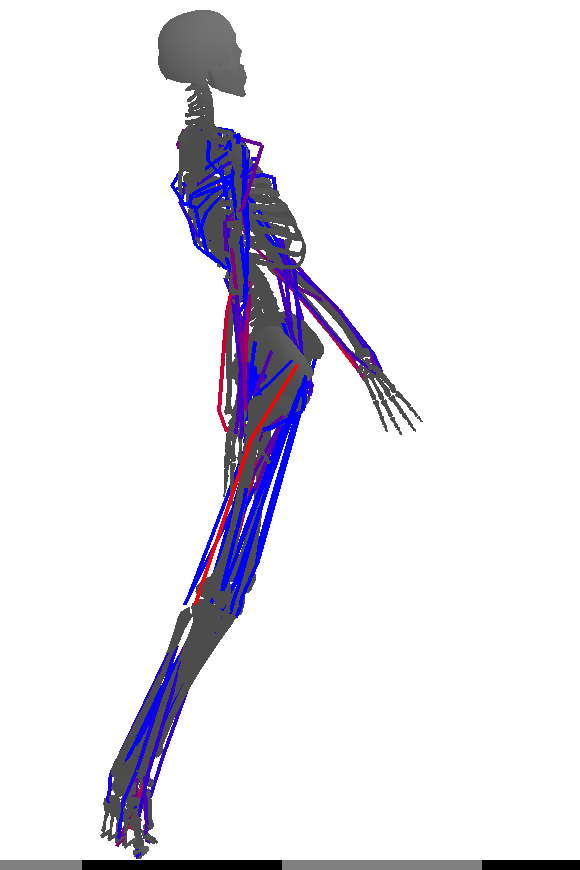}
  \includegraphics[width=0.1\linewidth]{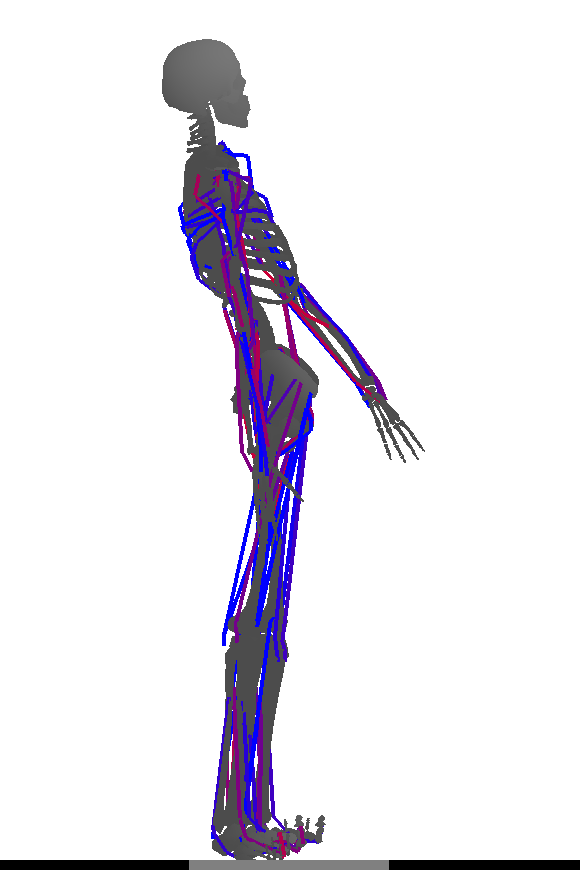}
  \includegraphics[width=0.1\linewidth]{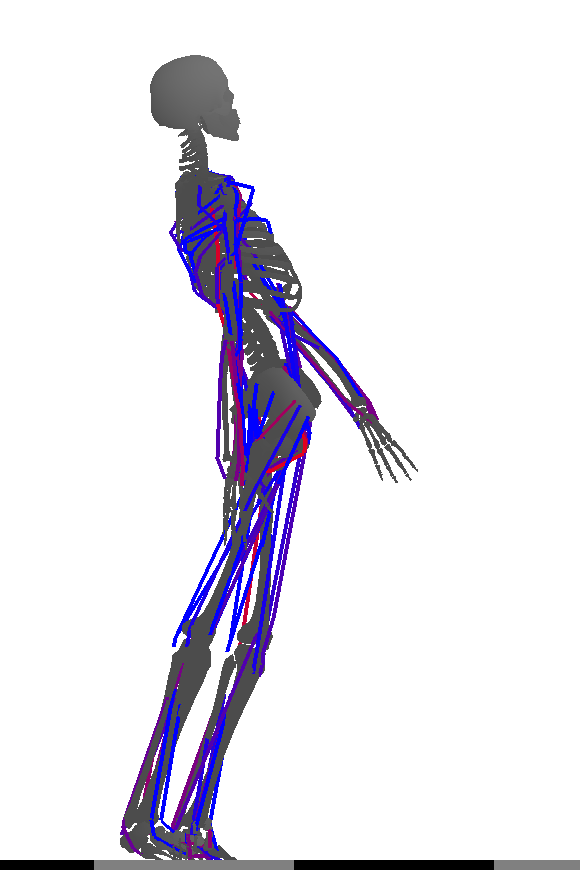} 
  \includegraphics[width=0.1\linewidth]{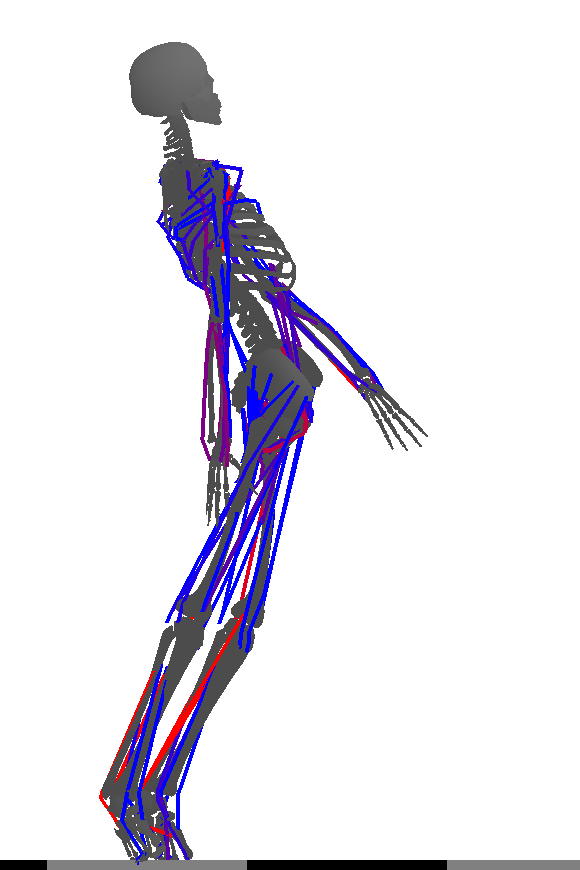} 
  \includegraphics[width=0.1\linewidth]{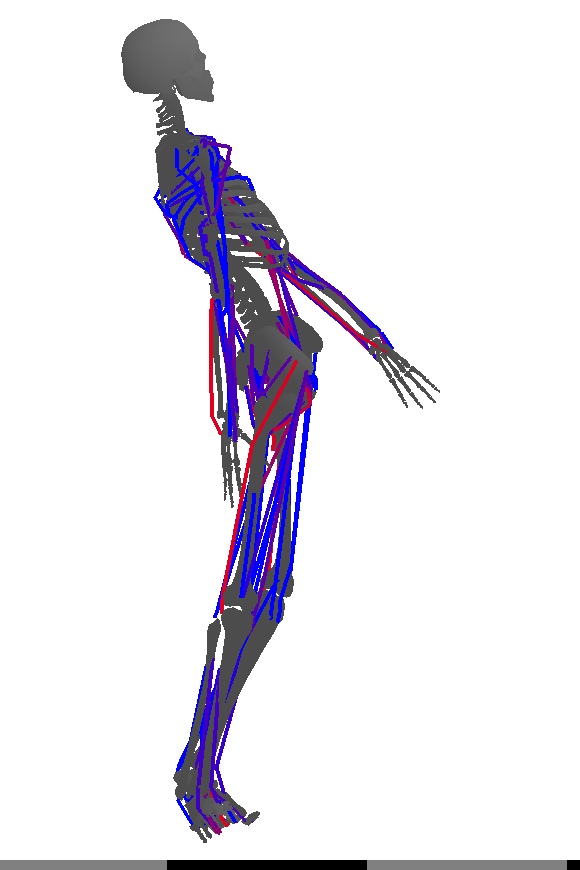}
  \includegraphics[width=0.1\linewidth]{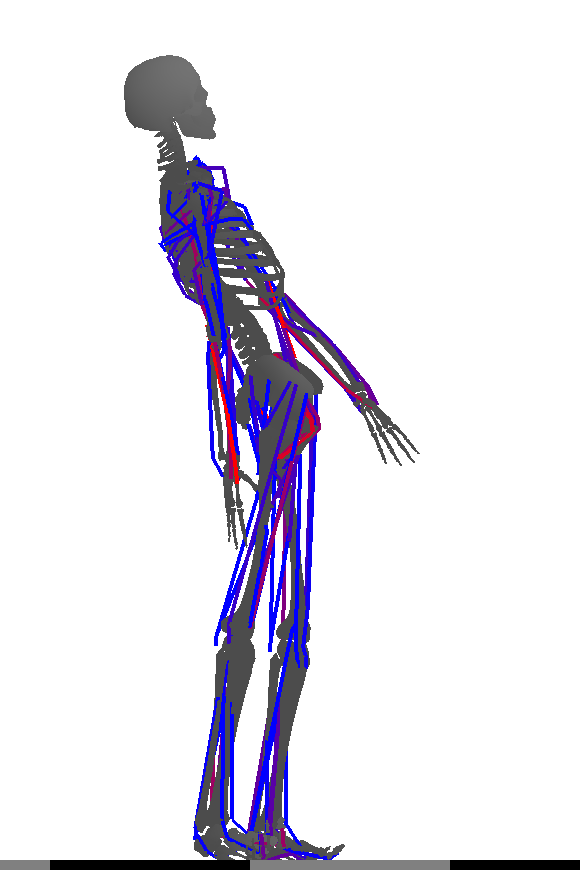} 
  \includegraphics[width=0.1\linewidth]{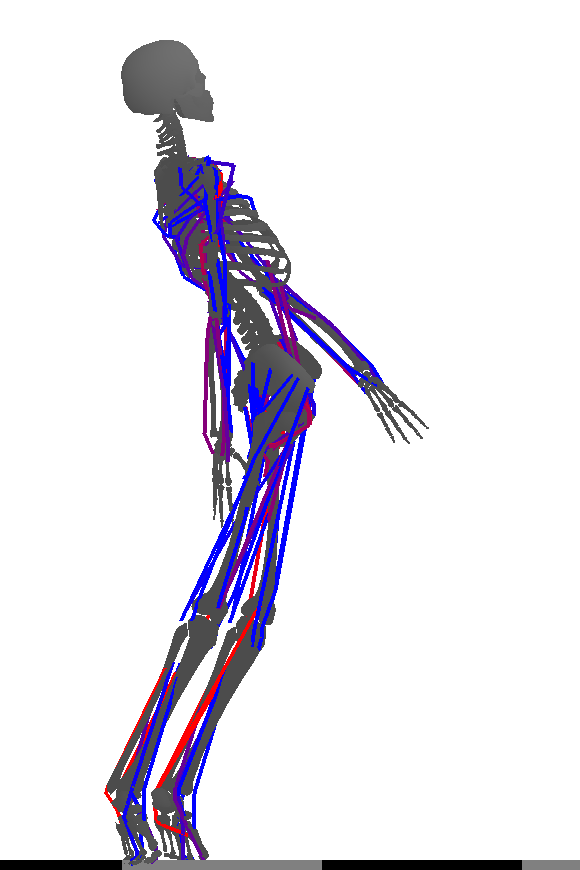}    
 
  \includegraphics[width=0.1\linewidth]{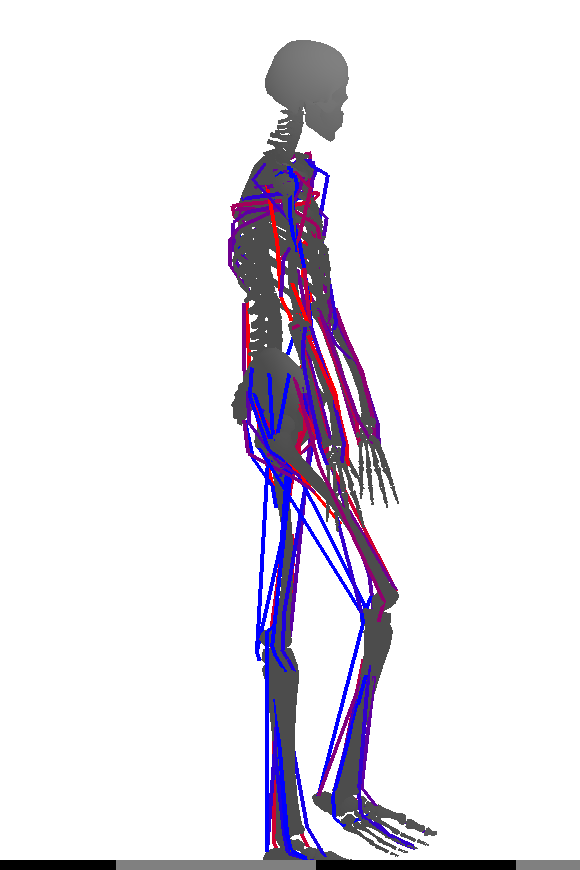}
  \includegraphics[width=0.1\linewidth]{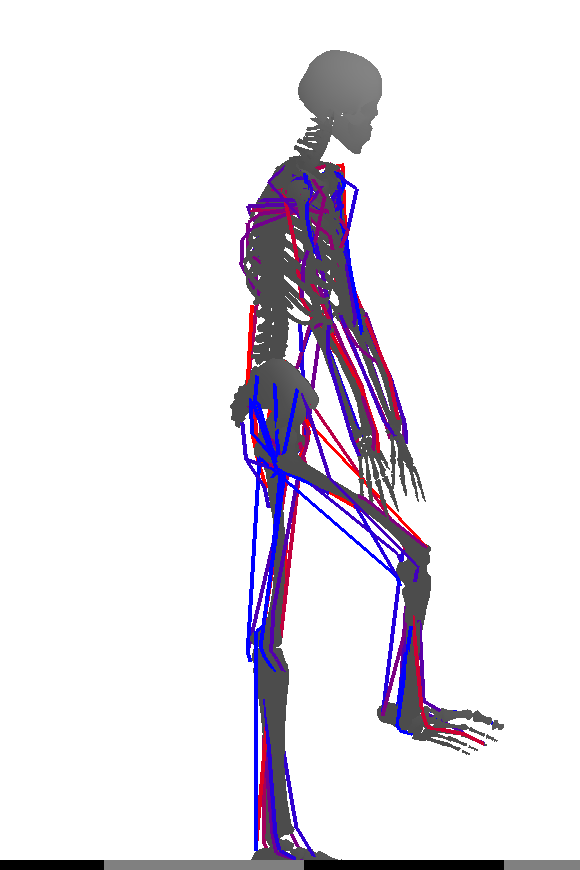} 
  \includegraphics[width=0.1\linewidth]{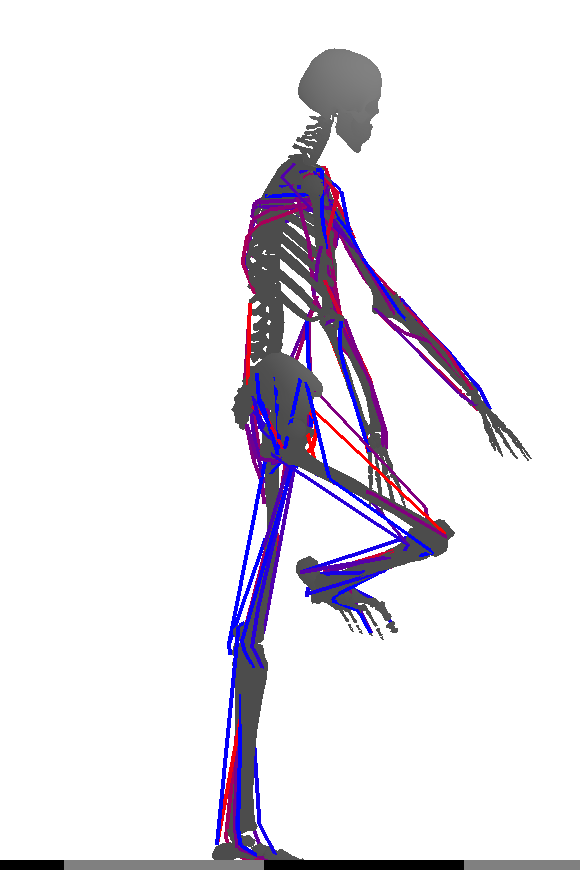}
  \includegraphics[width=0.1\linewidth]{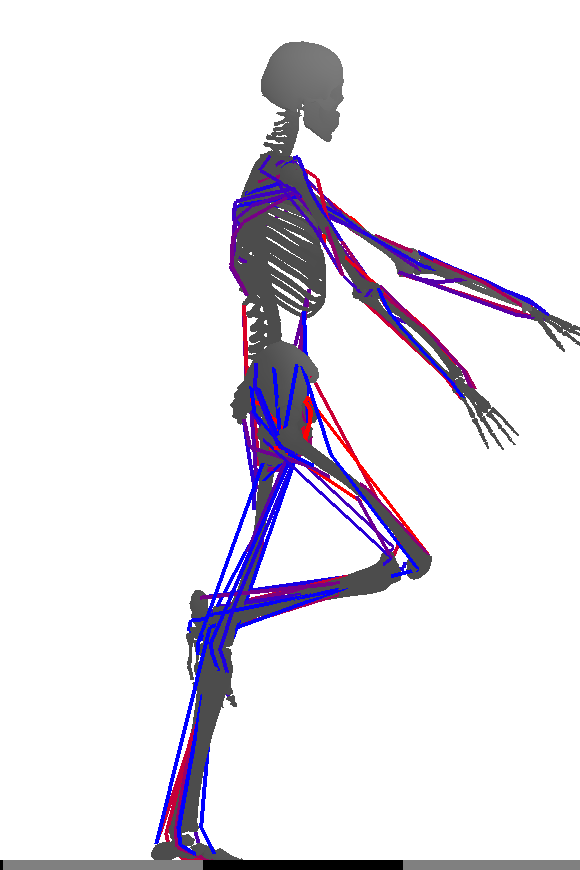}
  \includegraphics[width=0.1\linewidth]{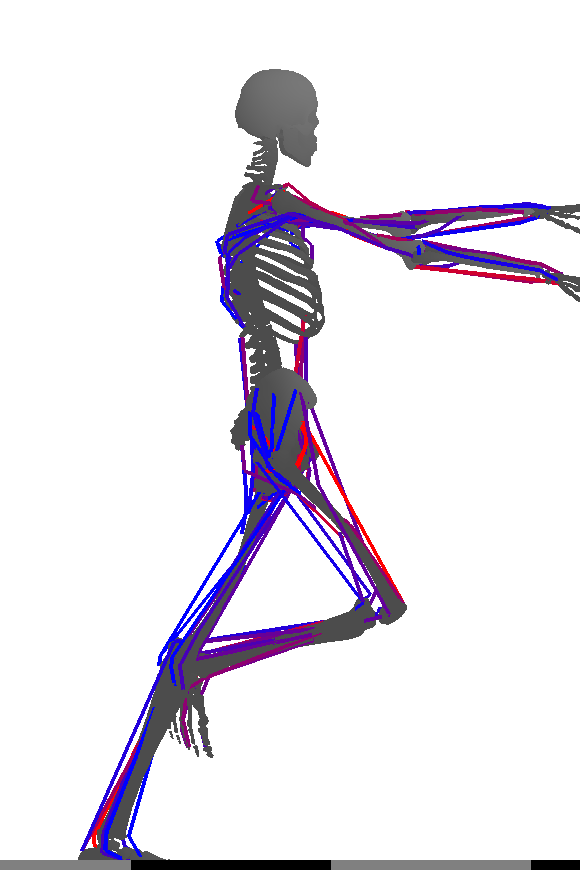} 
  \includegraphics[width=0.1\linewidth]{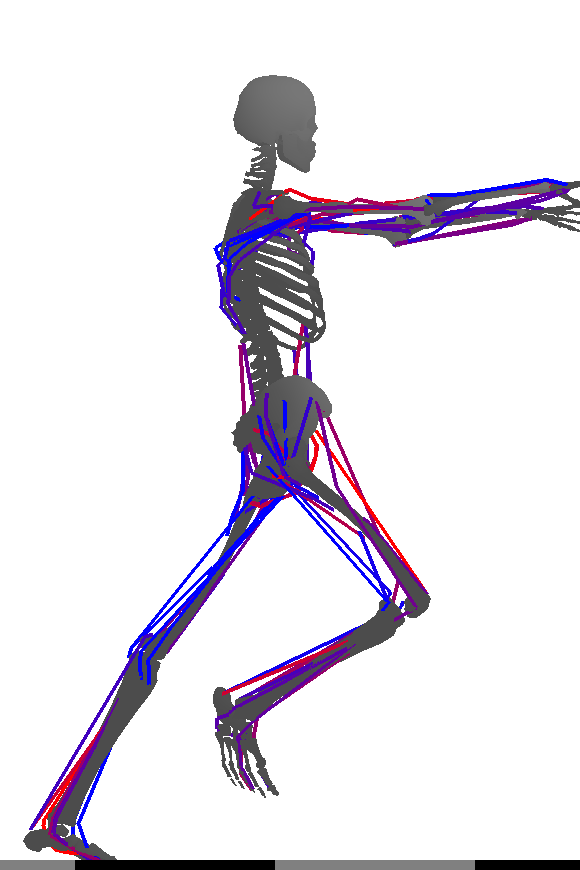} 
  \includegraphics[width=0.1\linewidth]{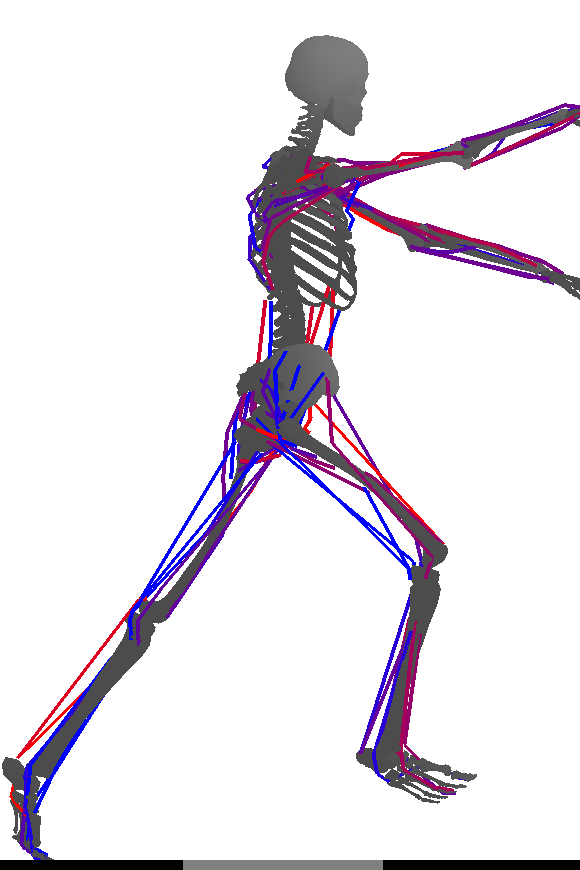}
  \includegraphics[width=0.1\linewidth]{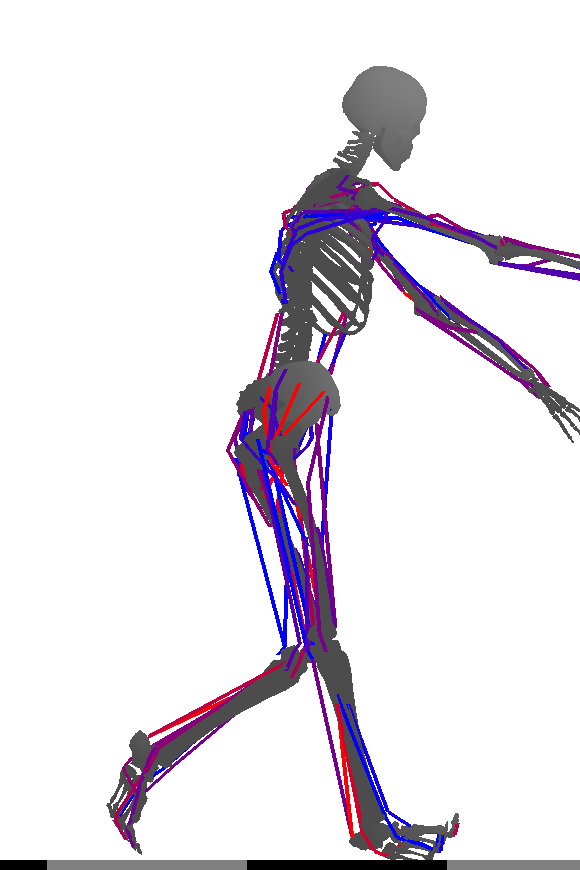} 
  \includegraphics[width=0.1\linewidth]{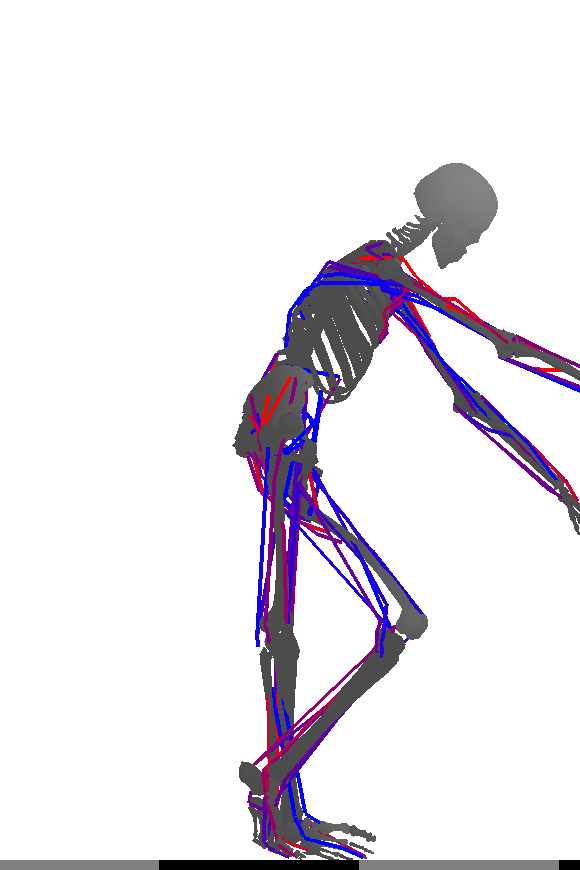}  
  
  \includegraphics[width=0.1\linewidth]{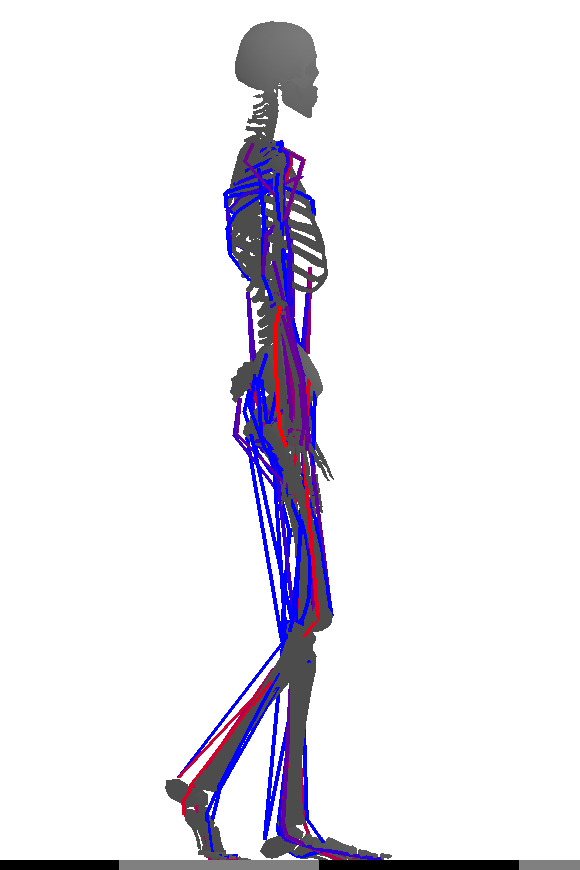}
  \includegraphics[width=0.1\linewidth]{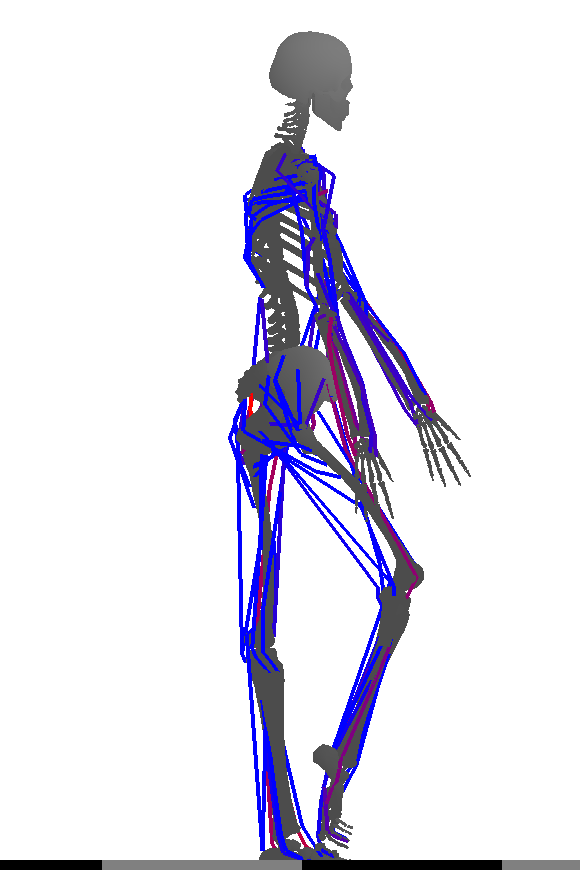} 
  \includegraphics[width=0.1\linewidth]{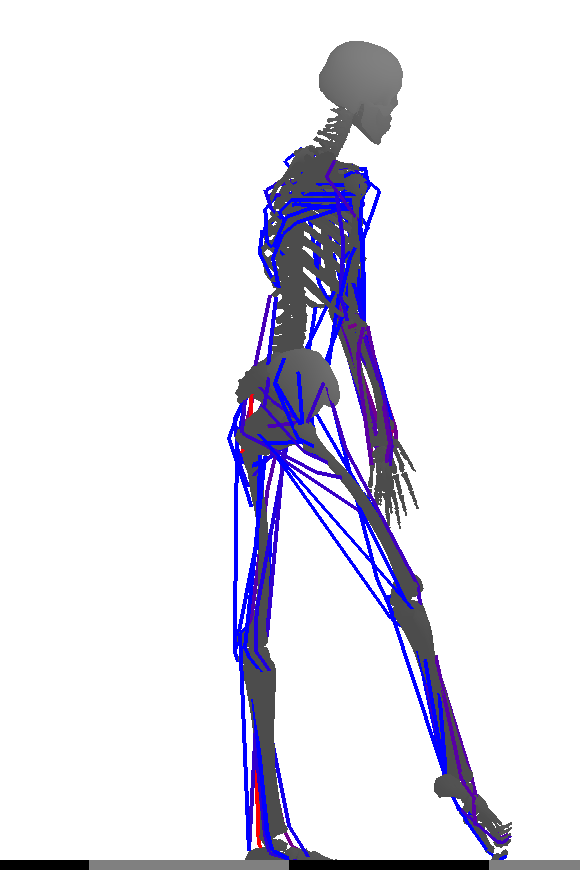}
  \includegraphics[width=0.1\linewidth]{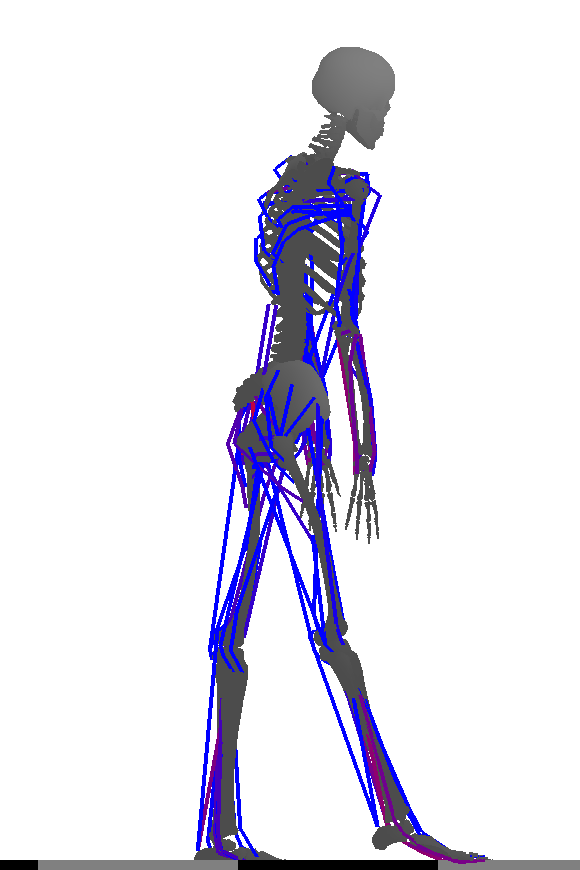}
  \includegraphics[width=0.1\linewidth]{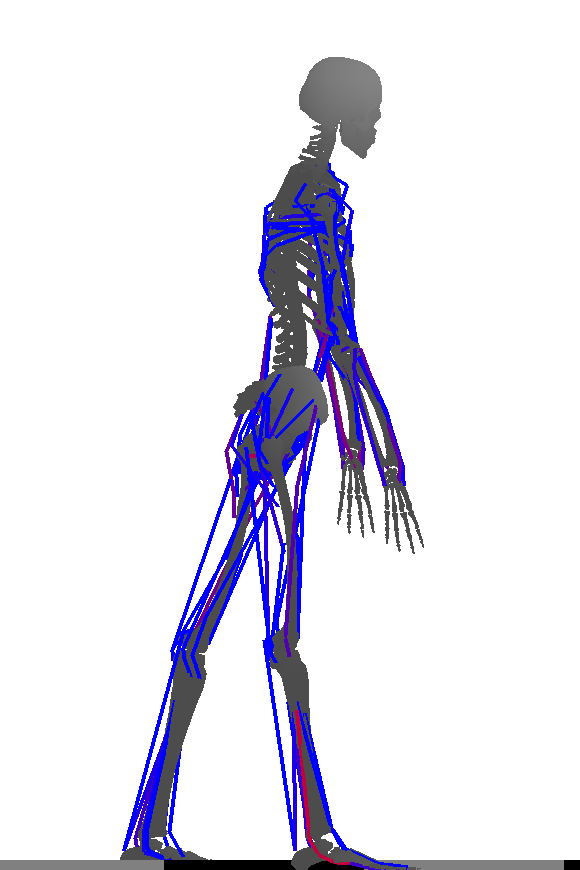} 
  \includegraphics[width=0.1\linewidth]{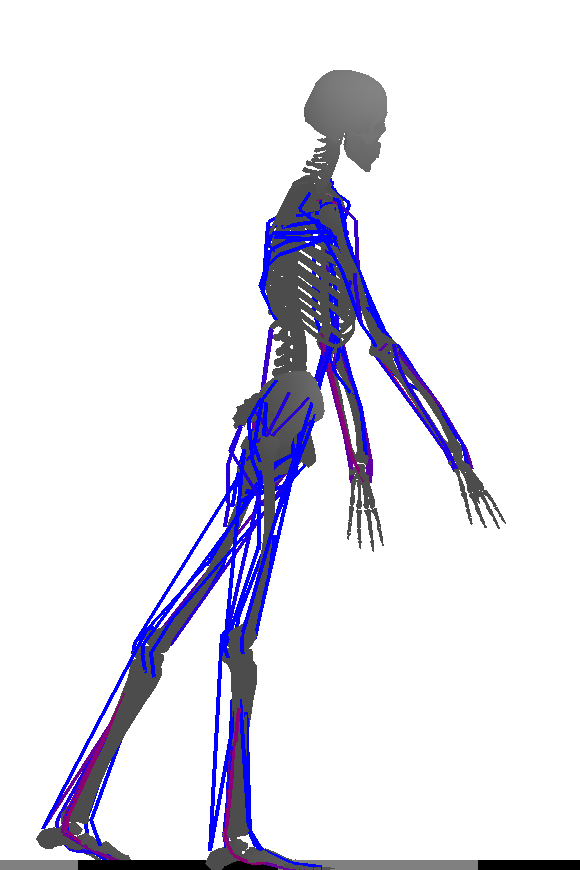} 
  \includegraphics[width=0.1\linewidth]{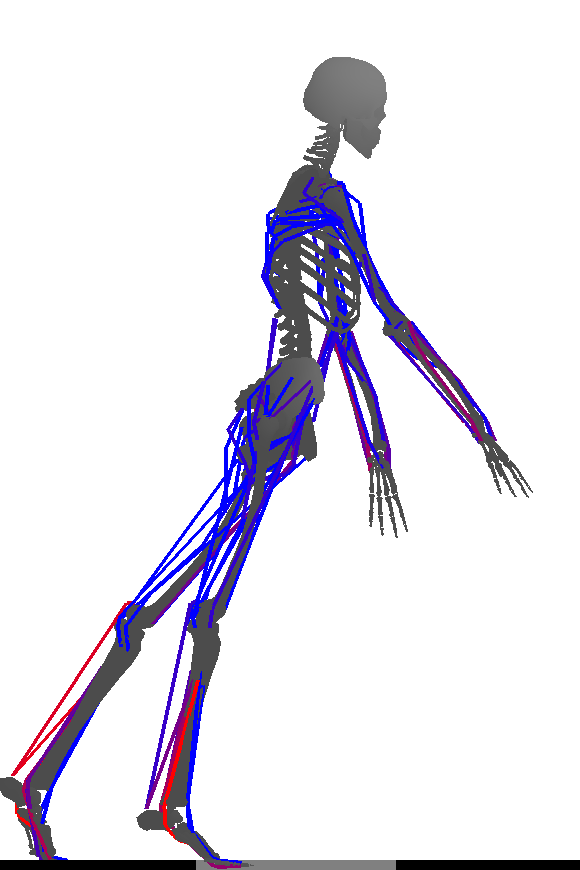}
  \includegraphics[width=0.1\linewidth]{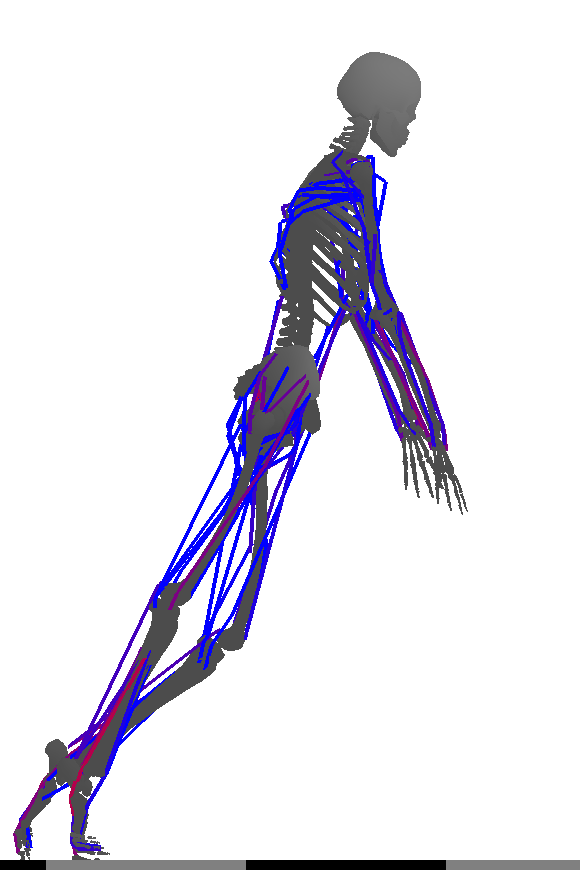} 
  \includegraphics[width=0.1\linewidth]{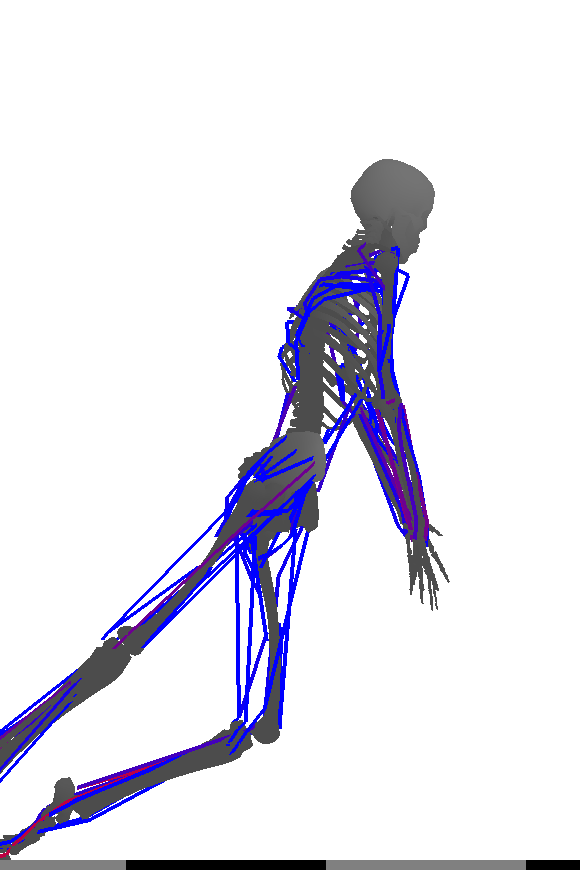}

  \caption{Experimental results. From top to bottom: Ours (row 1), Dense energy (MET) only (row 2), Sparse energy (CoT) only (row 3), Without energy reward (row 4), Start with a double stance pose (row 5), Dense activation reward (row 6), Without muscle fiber length in observation (row 7)}
\end{figure*}

\clearpage

\begin{figure}[h]
  \centering
  \includegraphics[trim={1cm, 2cm, 1cm, 1cm}, clip, width=1.\linewidth]{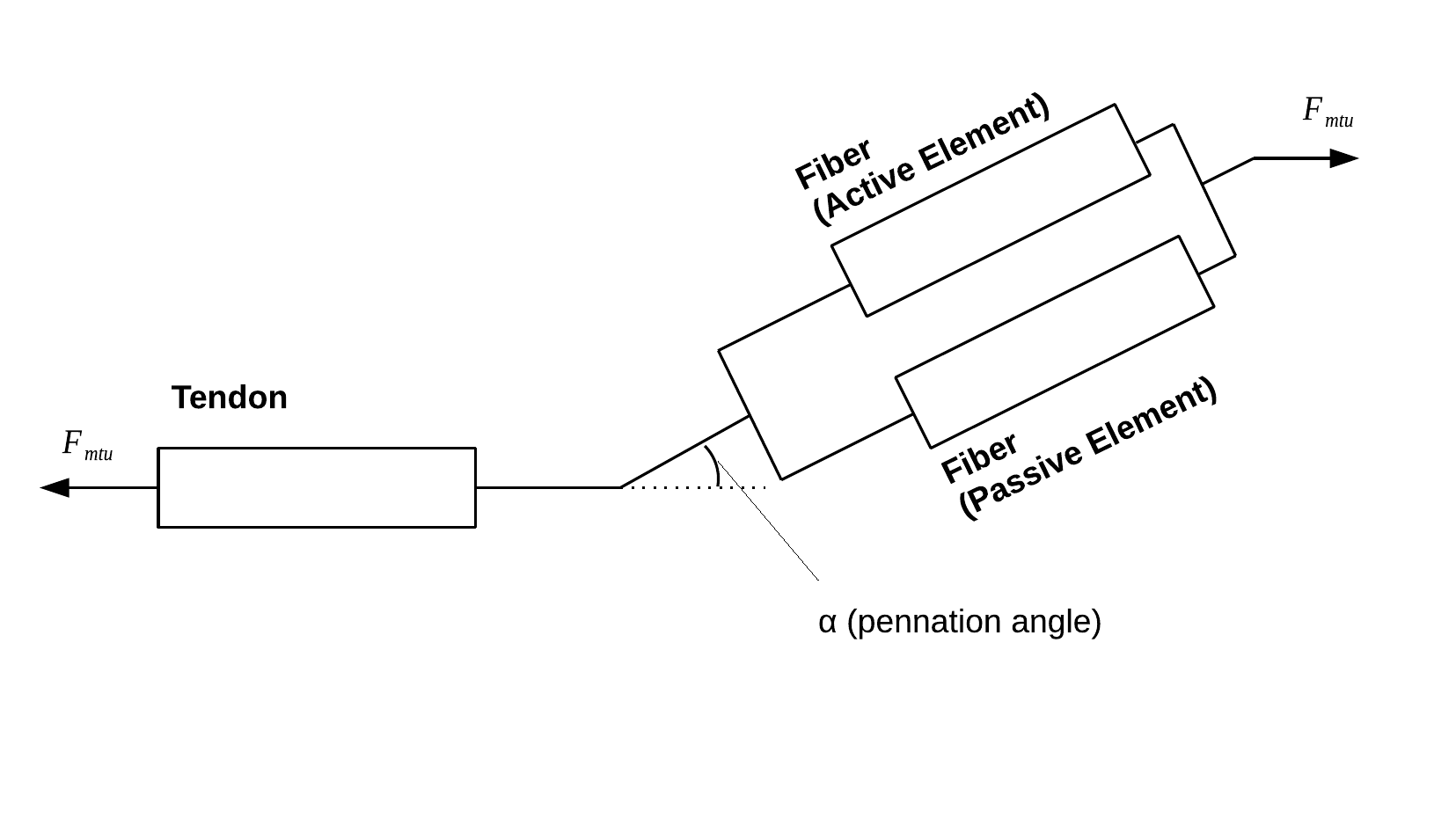} %
  \caption{Hill-type muscle model\\}
\end{figure}

\end{document}